\documentclass[11pt,a4paper]{article}
\usepackage{amsmath,amssymb}
\usepackage{jheppubx}
\usepackage{graphicx}
\usepackage{pdfpages}
\usepackage{booktabs}

\usepackage[usenames,dvipsnames]{xcolor}
\usepackage{cancel, ulem}
\usepackage{amsmath}

\input epsf
\usepackage{epstopdf}
\usepackage{ upgreek }

\definecolor{darkgreen}{rgb}{0,0.4,0}
\definecolor{darkred}{rgb}{0.4,0,0}
\definecolor{darkblue}{rgb}{0,0,0.4}
\definecolor{DarkGreen}{rgb}{0,0.5,0.3}

\def\be{\begin{equation}}
\def\ee{\end{equation}}
\newcommand{\bea}{\begin{eqnarray}}
\newcommand{\eea}{\end{eqnarray}}

\input epsf

\newlength{\extraspace}
\newlength{\extraspaces}
\def\a{\alpha}

\def\II{\relax{I\kern-.10em I}}

\def\IZ{\relax{\rm Z\kern-.34em Z}}
\def\IB{\relax{\rm I\kern-.18em B}}
\def\IC{{\relax\hbox{$\inbar\kern-.3em{\rm C}$}}}
\def\ID{\relax{\rm I\kern-.18em D}}
\def\IE{\relax{\rm I\kern-.18em E}}
\def\IF{\relax{\rm I\kern-.18em F}}
\def\IG{\relax\hbox{$\inbar\kern-.3em{\rm G}$}}
\def\IGa{\relax\hbox{${\rm I}\kern-.18em\Gamma$}}
\def\IH{\relax{\rm I\kern-.18em H}}
\def\II{\relax{\rm I\kern-.18em I}}
\def\IK{\relax{\rm I\kern-.18em K}}
\def\IP{\relax{\rm I\kern-.18em P}}

\def\inbar{\,\vrule height1.5ex width.4pt depth0pt}

\def\IR{\relax{\rm I\kern-.18em R}}

\def\ls{l_s}

\def\lp10{\ell_p^{10}}
\def\lp11{\ell_p^{11}}
\def\R11{R_{11}}

\def\frac#1#2{{#1 \over #2}}

\newdimen\tableauside\tableauside=1.0ex
\newdimen\tableaurule\tableaurule=0.4pt
\newdimen\tableaustep
\def\phantomhrule#1{\hbox{\vbox to0pt{\hrule height\tableaurule width#1\vss}}}
\def\phantomvrule#1{\vbox{\hbox to0pt{\vrule width\tableaurule height#1\hss}}}
\def\sqr{\vbox{%
  \phantomhrule\tableaustep
  \hbox{\phantomvrule\tableaustep\kern\tableaustep\phantomvrule\tableaustep}%
  \hbox{\vbox{\phantomhrule\tableauside}\kern-\tableaurule}}}
\def\squares#1{\hbox{\count0=#1\noindent\loop\sqr
  \advance\count0 by-1 \ifnum\count0>0\repeat}}
\def\tableau#1{\vcenter{\offinterlineskip
  \tableaustep=\tableauside\advance\tableaustep by-\tableaurule
  \kern\normallineskip\hbox
    {\kern\normallineskip\vbox
      {\gettableau#1 0 }%
     \kern\normallineskip\kern\tableaurule}%
  \kern\normallineskip\kern\tableaurule}}
\def\gettableau#1 {\ifnum#1=0\let\next=\null\else
  \squares{#1}\let\next=\gettableau\fi\next}

 \def\eqnn#1{\xdef #1{(\secsym\the\meqno)}\writedef{#1\leftbracket#1}%
 \global\advance\meqno by1\wrlabeL#1}
 \def\eqna#1{\xdef #1##1{\hbox{$(\secsym\the\meqno##1)$}}
 \writedef{#1\numbersign1\leftbracket#1{\numbersign1}}%
 \global\advance\meqno by1\wrlabeL{#1$\{\}$}}
 \def\eqn#1#2{\xdef #1{(\secsym\the\meqno)}\writedef{#1\leftbracket#1}%
 \global\advance\meqno by1$$#2\eqno#1\eqlabeL#1$$}

\global\newcount\itemno \global\itemno=0

\def\itemaut#1{\global\advance\itemno by1\noindent\item{\the\itemno.}#1}

\def\({\left(}
\def\){\right)}

\def\cc{{\bf c}}

\def\pp{{\bf p}}

\usepackage[yyyymmdd,hhmmss]{datetime}
\newif{\ifeq}           
\eqtrue                 
\def\question#1
{
\ifeq
\textcolor{darkblue}{#1}
\fi
}

\nocite{*}

\def\sqd{^2}
\def\ll{_}
\def\uu{^}
\def\hh{{1\over2}}
\def\cc{\,}

\def\bi{\begin{itemize}}
\def\ei{\end{itemize}}
\def\e{\epsilon}

\newenvironment{PurpleEnv}%
{\color{Purple}}%
{\color{Black}}
\newenvironment{BlueEnv}%
{\color{Blue}}%
{\color{Black}}
{\color{Red}}%
{\color{Black}}

\def\QQTColor{Sapphire}

\newenvironment{QQTEnv}%
{\color{\QQTColor}}%
{\color{Black}}

\def\EQQQTColor{Purple}

\newenvironment{EQQQTEnv}%
{\color{\EQQQTColor}}%
{\color{Black}}

\usepackage{physics}
\usepackage{mathtools}

\usepackage{hyperref}
\hypersetup{
     colorlinks   = true,
     citecolor    = magenta
}

\def\outt#1{{}}

\renewcommand{\ll}{_}

\def\muchlessthan{< \hskip-.05in <}
\def\muchgreaterthan{> \hskip-.05in >}
\def\muchgreaterthan{> \hskip-.05in >}

\newcommand\ba{\begin{eqnarray}}
\newcommand\ea{\end{eqnarray}}
\newcommand{\bbb}{\begin{eqnarray}\begin{array}{c}}
\newcommand{\bbl}[1]{\ba\label{#1}\begin{array}{c}{c}}
\newcommand{\eee}{\end{array}\end{eqnarray}}
\def\een#1{\label{#1} \eee}
\def\xxn#1{\een{#1}  \bbb}
\def\xxnn{\nonumber\xxx}
\def\eenn{\nonumber\eee}

\def\NewParagraph{\nnp}
\def\nnp{\NewParagraph}
\renewcommand{\IR}{{\mathbb{R}}}

\renewcommand{\thefigure}{\arabic{figure}} 

 \makeatletter
    
    \@addtoreset{equation}{section}
  \makeatother

\makeatletter
\g@addto@macro\bfseries{\boldmath}
\makeatother

\def\rdots{\redd{\circ\circ\circ}}

\def\D{\Delta}
\def\d{\delta}

\def\ggW{{\color{Salmon}{g}}}

\def\ggW{\ggW}

\def\rwa#1{\underline{\bf {#1}}}

\def\gg{\nabla}

\def\rdots{{\color{Red} \circ\circ\circ}}

\def\pp{\partial}
\def\gg{\NAMBLA}
\def\NAMBLA{\nabla}
\def\m{\mu}
\def\r{\rho}
\def\k{{\kappa}}
\def\g{\gamma}
\def\a{\alpha}
\def\b{\beta}
\def\m{\mu}

\def\s{\sigma}

\def\r{\rho}
\def\th{\theta}
\def\G{\Gamma}

\def\xxx{\eee\bbb}

\def\rr#1{(\ref{#1})}

\definecolor{DarkGreen}{rgb}{0,.64,0}
\definecolor{gunmetal}{rgb}{0.171875, 0.207031, 0.222656}
\definecolor{chartreuse}{rgb}{.49,.98,0}
\definecolor{amethyst}{rgb}{0.59375,0.398438,0.792969}
\definecolor{brownrust}{rgb}{0.6875, 0.316406, 0.242188}
\definecolor{Violet}{rgb}{0.5,0,1}
\definecolor{BurntOrange}{rgb}{0.792969,0.332031,0}
\definecolor{FreshEggplant}{rgb}{0.59375, 0., 0.414063}
\definecolor{salmon}{rgb}{0.996094,0.507813,0.410156}
 \definecolor{FrenchRose}{rgb}{0.96875, 0.292969, 0.5625}
\definecolor{Cabaret}{rgb}{0.808594, 0.242188, 0.46875}
\definecolor{Shamrock}{rgb}{0.242188, 0.808594, 0.582031}
\definecolor{RobinsEggBlue}{rgb}{0., 0.792969, 0.792969}
\definecolor{GuardsmanRed}{rgb}{0.792969, 0., 0.}
\definecolor{Sapphire}{rgb}{0.183594, 0.328125, 0.621094}
\definecolor{Sorbus}{rgb}{0.996094, 0.429688, 0.0273438}
\definecolor{Red}{rgb}{1,0,0}
\definecolor{Blue}{rgb}{0,0,1}
\definecolor{Green}{rgb}{0,1,0}
\definecolor{thistle3}{rgb}{0.800781, 0.707031, 0.800781}
\definecolor{thistle4}{rgb}{0.542969, 0.480469, 0.542969}

\definecolor{SHGColor}{rgb}{0.996094, 0.429688, 0.0273438}

{\color{SHGColor} \it }%
{--SH~~\rm\color{Black}}

\definecolor{Blue}{rgb}{0,0,1}

\def\co{{\cal O}}
\def\dag{^\dagger}
\def\pr{^\prime}

\def\llsk{\hskip.5in}

\def\st{^{*}}
\def\ls#1{_{[{{#1}}]}}
\def\lrm#1{_{\rm {#1}}}
\def\ups#1{^{[{{#1}}]}}
\def\uprm#1{^{(\rm {#1})}}
\def\lp#1{_{({{#1}})}}
\def\upp#1{^{({{#1}})}}

\def\xxx{\eee\bbb}

\def\k{\kappa}

 \definecolor{chartreuse}{rgb}{.49,.98,0}
\definecolor{amethyst}{rgb}{0.59375,0.398438,0.792969}
\definecolor{brownrust}{rgb}{0.6875, 0.316406, 0.242188}
\definecolor{BurntOrange}{rgb}{0.792969,0.332031,0}
\definecolor{FreshEggplant}{rgb}{0.59375, 0., 0.414063}
\definecolor{salmon}{rgb}{0.996094,0.507813,0.410156}
 \definecolor{FrenchRose}{rgb}{0.96875, 0.292969, 0.5625}
\definecolor{Cabaret}{rgb}{0.808594, 0.242188, 0.46875}
\definecolor{Shamrock}{rgb}{0.242188, 0.808594, 0.582031}
\definecolor{RobinsEggBlue}{rgb}{0., 0.792969, 0.792969}
\definecolor{GuardsmanRed}{rgb}{0.792969, 0., 0.}
\definecolor{Sapphire}{rgb}{0.183594, 0.328125, 0.621094}
\definecolor{Sorbus}{rgb}{0.996094, 0.429688, 0.0273438}
\definecolor{Red}{rgb}{1,0,0}
\definecolor{Blue}{rgb}{0,0,1}
\definecolor{Purple}{rgb}{0.808594, 0.242188, 0.46875}

\def\redlowdash{{\color{Red}{\rule[-0.5ex]{2pt}{0.4pt}}}}
\def\bluelowdash{{\color{Blue}{\rule[-0.5ex]{2pt}{0.4pt}}}}
\def\redmiddash{{\color{Red}{\rule[+0.5ex]{2pt}{0.4pt}}}}

\def\cute{{\lower3.5pt\hbox{\sixly
  \kern-.21pt \char58 \kern-.21pt }}}

\def\midcute{{\lower-1.0pt\hbox{\sixly
  \kern-.21pt \char58 \kern-.21pt }}}

  \def\lowcute{{\lower3.5pt\hbox{\sixly
  \kern-.21pt \char58 \kern-.21pt }}}
  
  \def\redmidcute{{\color{Red} \midcute}}
  
  \def\redlowcute{{\color{Red} \lowcute}}
   
    \def\bluelowcute{{\color{Blue} \lowcute}}
    
    \def\redX{{\color{Red} X}}

   \def\swave{\bgroup \markoverwith \midcute \ULon} 
  
  \def\redswave{\bgroup \markoverwith \redmidcute \ULon} 
  
  \def\reduline{\bgroup \markoverwith \redlowdash \ULon}
  
   \def\blueuline{\bgroup \markoverwith \bluelowdash \ULon}
  
   \def\reduwave{\bgroup \markoverwith \redlowcute \ULon}
   
   \def\blueuwave{\bgroup \markoverwith \bluelowcute \ULon}

  \def\redsout{\bgroup \markoverwith \redmiddash \ULon}
  
   \def\bluesout{\bgroup \markoverwith \bluemiddash \ULon}

\def\redXout{\bgroup \markoverwith \redX \ULon}

    \def\rwa{\reduwave}

\def\o{\omega}
\def\O{\Omega}
\def\L{\Lambda}

\makeatletter
\newcommand*{\Relbarfill@}{\arrowfill@\Relbar\Relbar\Relbar}
\newcommand*{\xeq}[2][]{\ext@arrow 0055\Relbarfill@{#1}{#2}}
\makeatother

\def\area#1{{\cal A}\ll{({#1})}}
\def\prpr{^{\prime\prime}}
\def\prprpr{^{\prime\prime\prime}}

\def\chh{\hat{\chi}}

\def\emm#1{{\it {#1}}}

\def\apr{{\alpha^\prime}}

\def\k{\kappa}

\def\bii{\begin{itemize}}
\def\ei{\end{itemize}}

\def\SHISHereMacro{{\rm here}}
\def\FORMacro{ {{\rm Favrod-Orlando-}\atop{\rm Reffert}} }
\def\KPMacroA{{{\rm Kravec-}\atop{\rm Pal}}}
\def\SWMacroA{{{\rm Son-}\atop{\rm Wingate}}}

\def\emm#1{\rwa{#1}}
\def\bbd#1{${#1}$}
\begin{document}

\preprint{IPMU20-0107}
\title{Droplet-Edge Operators in Nonrelativistic Conformal Field Theories}
\author{Simeon Hellerman$^1$ \&}
\author{Ian Swanson$^2$}
\affiliation{\it $^1$Kavli Institute for the Physics and Mathematics of the Universe \textsc{(wpi)}\\
The University of Tokyo \\
 Kashiwa, Chiba  277-8582, Japan\\
}
\emailAdd{$^1$simeon.hellerman.1@gmail.com}
\emailAdd{$^2$ianswanson.physics@gmail.com}

\abstract{
We consider the large-charge expansion of the charged ground state of a Schr\"odinger-invariant, nonrelativistic
conformal field theory in a harmonic trap, in general dimension $d$.  In the existing literature, 
the energy in the trap has been computed to next-to-leading order (NLO) at large charge $Q$, 
which comes from the classical contribution of two higher-derivative terms in the effective field theory.  In
this note, we explain the structure of operators localized at the edge of the droplet, where the density drops to zero.  We 
list all operators contributing to the ground-state energy with nonnegative powers of $Q$ in the large-$Q$ expansion.  
As a test, we use dimensional 
regularization to reproduce the calculation of the NLO ground state energy by Kravec and Pal  \cite{Kravec:2018qnu}, and we recover 
the same universal coefficient for the logarithmic term
as in that work.  We refine the derivation by presenting a systematic operator analysis of the 
possible edge counterterms, showing that different choices of cutoff procedures must yield the same renormalized result 
up to an enumerable list of Wilson coefficients for conformally invariant local counterterms at the droplet edge.  
We also demonstrate the existence of a previously unnoticed edge contribution to the ground-state operator dimension 
of order $ Q\uu{{2\over 3} - {1\over d}}$ in $d$ spatial dimensions.
Finally, we show there is no bulk or edge counterterm scaling as $Q\uu 0$ in two spatial dimensions, which establishes the 
universality of the order $Q\uu 0$ term in large-$Q$ expansion of the lowest charged operator dimension in $d=2$.}
\maketitle

\section{Introduction and Summary}

Various zero-temperature phase transitions are purported to be described by ``quantum critical behavior," defined by
a nonrelativistic conformal field theory.   In this paper, we deal with the NRCFT studied by Son and Wingate
\cite{Son:2005rv} that describes interacting fermions at zero temperature when a parameter 
of the Hamiltonian has been tuned (in all existing cases,
by dialing the strength of a background magnetic field near a Feshbach resonance; 
see, e.g.,~\cite{OHara:2002pqs,OHara:2002oaq} for early experimental realizations) so 
that the fermionic scattering length becomes 
infinite.  This notional fixed point, originally posed by Bertsch \cite{Bishop:2001rf,Baker:1999np}, 
is referred to in the literature as a {\it unitary} fermi gas.  

In the case of a trapping potential where the charge is supported in a finite region, the precision of the computation of the ground-state
energy is limited by one's understanding of the nature of the Hamiltonian density near the edge of the atomic droplet, where the 
charge density goes to zero.  

To make this slightly more explicit, let us briefly introduce some basics of the Son-Wingate NRCFT.
The symmetries of the theory include temporal- and spatial-translation symmetry, atom-number (mass) conservation,
Galilean relativistic invariance, and scale invariance, which is enhanced to a nonrelativistic conformal
symmetry.  The full group generated by these symmetries is known as the Schr\"odinger group.  This symmetry 
is respected by free, nonrelativistic fermions in any spatial dimension $d$ (but there are believed to be
non-free examples realized in nature).
The condensate droplet effective theory descends from a general treatment of massive
{Schr\"odinger} particles $\psi$, coupled to an external potential consistent with the symmetries
of the system.  Invariance under a phase rotation of $\psi$ is a global $U(1)$ symmetry.
At infinite scattering length, the remaining physical degree of freedom in the theory 
is the phase of the condensate; for bosonic particles this phase is that of the field $\psi$ in the usual way, 
while for fermions one can identify it with (half) the phase of a Cooper pair.
The global $U(1)$ symmetry is broken by the choice of ground state, which fixes the charge $Q$, 
and fluctuations around this ground state are (conformal) Goldstone fields $\chi$;
the classical superfluid ground state with chemical potential $\mu$ is at $\chi = \mu\, t$.
The expansion of the EFT in the region of finite density is an expansion in derivatives of
$\chi$, where the conformal dimension is adjusted to marginality by fractional negative powers of 
\be
{\bf X} \equiv \dot{\chi} - A\ll 0 - {1\over{2m}}\cc (\vec{\pp}\chi - \vec{A})\sqd\ ,
\ee
and where the background potential has $\vec{A} = 0$ for the case of the harmonic trap.
The density $\r$ goes as $(m\cc {\bf X})\uu{{d\over 2}}$ in spatial dimension $d$, 
so the expansion is in powers of $\r\uu{-{1\over d}} \cc \vec{\pp}$, 
which breaks down at the droplet edge where the density falls to zero.

The present state of understanding was advanced by \cite{Favrod:2018xov, Kravec:2018qnu}, wherein 
the large-charge expansion of the NRCFT was presented as a way to perturbatively suppress quantum effects below
leading-order contributions in the Lagrangian.  
The authors of \cite{Kravec:2018qnu}, in particular, studied the system in a harmonic potential
trap, though the quantum effects of the theory remain  
uncontrolled in the absence of a detailed and complete
treatment of the structure of counterterms beyond those supported in the bulk of the density distribution.

More specifically, the breakdown of the derivative expansion in the interior is associated with unknown
contributions to the operator dimension, scaling with positive powers of the charge $Q$.  These unknown contributions are
parametrically larger than contributions from quantum-mechanical fluctuations of the $\chi$ field, which start
with the Casimir energy at order $Q\uu 0$.  Without the ability to test these quantum-mechanical contributions
against experiment, or against other methods of calculation, it remains open
whether the Lagrangian of \cite{Son:2005rv} is a true effective theory of the large-charge sector, 
or whether there are other light degrees of freedom at large $Q$ in addition to the conformal Goldstone field $\chi$.  
It is thus crucial to understand the contributions to the energy from singularities near the droplet edge, where the density falls to zero,
if we are to probe the quantum-mechanical completeness of the $\chi$ theory as a description of the low-energy states of the 
large-charge Hilbert space.
In this paper we aim to do precisely this.  We will explain the general structure of the droplet-edge terms and demonstrate
their role in the renormalization of the Hamiltonian at the classical and quantum level.


In spite of the singularity of the classical solution of the EFT near the droplet edge, the edge effects are in fact under control within
the EFT itself (as we will show), with the singular behavior being parametrized by new Wilson coefficients for additional effective terms localized
at the droplet edge.  To put the analysis in context, a similar situation occurs in the case of the large-spin expansion
of the relativistic effective string with freely moving endpoints \cite{Hellerman:2013kba}.  Here, the length of the string
goes as $R\lrm{phys}  = E\lrm{IR}\uu{-1} =  \sqrt{J\apr}$, and the local strong-coupling scale in the interior of the worldsheet
is the total energy of the string $E\lrm{UV}\uprm{interior} = \sqrt{J/{\apr}}$, while the effective loop-counting
parameter is $E\lrm{IR}  / E\lrm{UV} = {1 / J}$.  
The classical solution controlling leading-order observables is 
singular at the boundary of the worldsheet, and the large-spin expansion 
appears to break down entirely.  On closer examination \cite{Hellerman:2016hnf}, however,
one finds this is not so.  At the boundary of the worldsheet, the local strong coupling scale drops 
to $E\lrm{UV}\uprm{boundary} = J\uu{-{1\over 4}} \cc \apr\uu{-\hh}$.  Quantum
effects near the boundary are not as strongly suppressed as in the interior, but they are still suppressed: The
effective loop-counting parameter near the boundary is $J\uu{-{1\over 4}}$ rather than $J\uu{-1}$.

The breakdown of the bulk derivative expansion near the droplet edge of the NRCFT is resolved in precisely the same way as in the case
of the effective string.  At the droplet edge, the field appearing in fractional or negative powers in the
derivative expansion (which we refer to as the ``dressing field," by analogy with the case of the effective string) is 
\bbb
{\bf Y} / m \equiv (\vec{\pp} {\bf X})\sqd / m\ ,
\een{YDefPrecap}
instead of ${\bf X}$.  The infrared energy scale in the harmonic trap is
simply the trapping frequency $\o$, and the local strong-coupling energy 
scale, by way of the chemical potential $\m$, is 
$\langle({\bf Y} / m)\uu{{1\over 3}}\rangle \propto (\o\sqd \m)\uu{{1\over 3}} \sim Q\uu{{1\over{3d}}}\cc \o$,
instead of $\m\sim Q\uu{{1\over d}}\cc \o$.  (Here we use
the proportionality $Q \propto \m\uu d$ between the
chemical potential and total charge in the harmonic potential, as reviewed
in Sec.~\ref{LeadingOrderRelationsReview}.)  It follows that higher derivative terms at the edge are suppressed by powers of
$E\lrm{UV}\uprm{edge} / E\lrm{IR} \sim (\m / \o)\uu{{1\over 3}} \sim Q\uu{{1\over{3d}}}$
rather than powers of the hierarchy
$E\lrm{UV}\uprm{interior} / E\lrm{IR} \sim Q\uu{{1\over d}}$, which suppresses quantum effects and higher-derivative 
terms in the bulk.


The plan of the paper is as follows:
\bi
\item{In Section \ref{SonWingateReviewSection} we review the basic setup of the theory, including leading-order and next-to-leading-order (NLO) effects
in homogeneous ground states in the harmonic trap in $d$ dimensions, as calculated previously
in, e.g., ~\cite{Son:2005rv, Kravec:2018qnu}.}
\item{In Section \ref{DropletEdgeOps} we systematize the construction of edge counterterms and give rules for counting the $\m$-scaling of their contribution to the ground state energy in the harmonic potential.}
\item{In Section~\ref{NLOTrapEnergies} we calculate the contributions from NLO operators in the bulk using dimensional regularization, reproducing
the results of \cite{Kravec:2018qnu} up to nonlogarithmic contributions at order $Q\uu\hh$, 
corresponding to the (now elucidated) conformally invariant counterterms at the droplet edge. }
\item{In Section~\ref{conclusions} we discuss the results and summarize our conclusions.}
\ei


\section{The interacting unitary fermion NRCFT and its large-charge universality class}\label{SonWingateReviewSection}

Let us begin by reviewing the structure of the effective Lagrangian of \cite{Son:2005rv}.  We will keep our conventions
as close as possible to those of Son and Wingate, deviating only when necessary to extend those conventions in a natural way to arbitrary
complex spatial dimension $d$, or to make contact with alternate conventions used in recent literature, such as \cite{Kravec:2018qnu}.

\subsection{NRCFT: General structure}

The best-known interacting NRCFT is the theory of interacting fermions at the so-called unitary limit, where the scattering
length becomes infinite.  Femions at unitarity, unlike free fermions, are an example
of a system with no additional internal symmetries. Collections of various types of 
fermionic atoms \cite{Regal:2004zza,Zwierlein:2004zz, Kaplan:1998tg,Kaplan:1998we, Chin:2001uan,Roberts:1998zz} 
are all believed to flow to the same critical point at zero
temperature. 

The interacting critical point can be distinguished easily from the free critical point, for instance, by the ground state energy density
for a homogeneous state of fermion density $\r$.  
In general complex spatial dimension $d$, however, there is no natural choice for the number of spin states, and 
we need to exercise caution in expressing the free energy density if we wish to make contact 
with the conventions of \cite{Nishida:2010tm}, for instance.  
In \cite{Nishida:2010tm}, the authors normalize the ground state energy of the system relative to two species of scalar 
fermions in spatial dimension $d$, rather than a spinorial $SO(d)$ multiplet of 
spinning fermions.\footnote{For more detail, see the formula given below eqn.~{(18)} of \cite{Nishida:2010tm} for the 
Fermi momentum $k\lrm F$ in terms of the density of the free fermion system in $d$ dimensions: The authors write 
{$k\lrm F = [ 2\uu{d-1} \pi\uu{d/2} \G({d\over 2} + 1)\cc n]\uu{{1\over d}}$}, with {$n$} denoting
the ground-state fermion density (we use $\r$).  Formula {(18)} of \cite{Nishida:2010tm} is correct for a system of free 
fermions with exactly two fermion states per momentum level, independent of
the dimension $d$.  For a different number $a\ll s$ of fermion states at a given momentum, the relationship 
would be $k\lrm F = [ a\ll s\uu{-1}\cc 2\uu d \pi\uu{d/2} \G({d\over 2} + 1) \cc n]\uu{{1\over d}}$. 
We discuss this further in Sec.~\ref{ConventionsForXiDefinitionSummary}.} 
The ratio of the ground state energy density of the interacting theory (which we shall refer to as the 
Son-Wingate (\textsc{sw}) theory) to that of the corresponding 
free theory with two scalar particle species at fixed $\r$
is thus explicitly stated:
\bbb
\xi \equiv {{{\cal H}\lrm{interacting\, NRCFT}}\over{{{\cal H}\lrm{free~fermion;\ 2\, scalar\, species}}}} \cc \bigg |\lrm{same~\r}\ .
\een{bertschxi}

Known as the Bertsch parameter, 
$\xi$ constitutes a characteristic dimensionless number of the critical point.
Straightforwardly, if $\xi$ is not equal to $1$, the critical point
does not describe the free fermion system.  
The parameter is particular to the NRCFT in question, 
analogous to the $c\ll{{3/2}}$ parameter of the large-charge relativistic conformal EFT in $2+1$ 
dimensions\cite{Hellerman:2015nra, Monin:2016jmo, Cuomo:2017vzg, Sharon:2020mjs, Gaume:2020bmp}, or its higher-dimensional \cite{Cuomo:2020rgt, Orlando:2019hte, Cuomo:2019ejv, Orlando:2020yii, Gaume:2020bmp} and nonabelian-symmetric \cite{Alvarez-Gaume:2016vff, Gaume:2020bmp, Hellerman:2017efx, Hellerman:2018sjf} counterparts.
The value of $\xi$ is approximately the same for all experimental realizations of the critical theory referred to above
(and different from unity, of course);  it can also be extracted numerically via Monte Carlo simulation \cite{Carlson:2003zz, Astrakharchik:2004zz}, 
placing its estimated value in three spatial dimensions around
\bbb
\xi\lrm{d=3} \simeq 0.4\ .
\eee

\subsection{EFT description of the large-charge sector}

As it stands, the definition of this NRCFT is unclear, a priori.  
As with the Wilson-Fisher $O(2)$ model \cite{Wilson:1971dc}, 
interacting NRCFTs are generically strongly coupled with 
non-infinitesimal anomalous dimensions, and any Lagrangian description would necessarily be
of Wilsonian type, with an infinite number of higher-derivative terms for whatever field content, 
all of them comparable in size (in units of the cutoff).
Outside a perturbative treatment of the theory near a weak-coupling 
region (such as an $\e$-expansion, as in \cite{Wilson:1971dc}, or the large-$N$ expansion), there is currently no
known way to verify or exclude the existence of a renormalization-group fixed point with a given set 
symmetries and degrees of freedom.  

Even assuming the existence of a Wilsonian description, such a treatment is of limited direct utility for the computation
of many observables of interest.  For certain limits in observable-space, however, the Wilsonian description of the fixed point becomes
effectively weakly coupled.  In the case of generic {\it relativistic} CFT, the properties of the ground state at charge $Q$ can be computed
in an asymptotic expansion in inverse powers of $Q$ 
\cite{Hellerman:2013kba, Hellerman:2015nra, Monin:2016jmo, Alvarez-Gaume:2016vff, Hellerman:2017veg, Hellerman:2017sur, Hellerman:2018xpi,Hellerman:2017efx,Cuomo:2017vzg,
Polchinski:1991ax, Aharony:2009gg,Aharony:2010cx,Hellerman:2016hnf,Aharony:2010db,Aharony:2011ga,Aharony:2011gb,Banerjee:2017fcx}.   
In examples with $U(1)$ symmetry, this large-charge sector is described by a conformally invariant effective Lagrangian for a single 
Goldstone boson $\chi$; in nonabelian and supersymmetric examples, the effective theories are described by 
generalizations that extend to the minimal field content as dictated by the symmetries.

The proposal in \cite{Son:2005rv} comprises an analogous conformally invariant Lagrangian in the {\it nonrelativistic} case,
describing the quantum critical points in interacting fermion systems at unitarity.
The theory is purported to control the dynamics of the large-charge limit of the NRCFT and,  
as in the aforementioned cases in relativistic CFTs, the large-charge sector admits an EFT for
a single conformal goldstone mode $\chi$, with a controlled derivative
expansion wherein higher-order terms make parametrically suppressed contributions at large $Q$.
Various leading-order and next-to-leading-order quantities have been computed  
for homogeneous ground states in vanishing background fields in infinite volume \cite{Son:2005rv,Nishida:2007pj},
in finite volume in flat space \cite{Favrod:2018xov}, and in a harmonic trapping potential \cite{Son:2005rv,Kravec:2018qnu}.  The energy in
the latter case is of particular importance, due to the nature of the state-operator correspondence (more on this below).

Note that this EFT may actually describe the large-charge dynamics of more than one critical point with the same
symmetries.  There may be other NRCFTs described by the same EFT with different values of $\xi$, as well as other subleading
Wilson coefficients, but the particular critical point describing the unitary Fermi gas stands as a 
useful testbed as a highly generic, nonempty nonrelativistic CFT.


\subsection{Relation to the relativistic state-operator correspondence}\label{StateOpCorrNRCFT}

As noted, there is an analogous large-charge EFT for the 
Wilson-Fisher $O(2)$ model \cite{Hellerman:2015nra} \cite{Monin:2016jmo}, 
and in fact the Son-Wingate EFT works out very similarly.
The difference most relevant in the present context is that
the state-operator correspondence in NRCFT is not between arbitrary local operators and states on the sphere, 
but between positively-charged local operators only, and states in flat space in a harmonic potential.  
States of zero or negative particle number do not correspond to any quantum state under the NRCFT
state-operator correspondence.

Specifically, for positively charged operators of charge $Q$, the corresponding
state is a state of charge $Q$ in infinite volume, with a nontrivial background potential 
\bbb
A\ll 0(\vec{x}) = {{m\o\sqd}\over 2} \vec{x}\sqd\ .
\een{HarmonicTrappingPotentialDef}
The scaling dimensions of charged operators in the NRCFT are equal to $\o\uu{-1}$ times the energies of the corresponding
states in the harmonic potential.

This is conceptually similar the $O(2)$ model, but the salient difference is that, while
the charged ground state of the $O(2)$ model on the sphere is spatially
homogeneous \cite{Hellerman:2015nra}, the charged ground state in the harmonic trap
is {\it inhomogeneous}, with the density falling to zero at some radius defining the finite extent of the droplet.  
The singularity at the edge of the droplet is resolved by unknown short-distance physics.
For purposes of low-energy, long-wavelength observables, the effects of 
these unknown dynamics can be absorbed into effective terms in the 
Hamiltonian that are localized at the surface of the droplet. Unlike
the case in a translationally invariant ground state, however, the singularity contributes
at the classical level; the Hamiltonian requires
regularization and renormalization even for tree-level consistency.  The analysis of
the classical UV singularity and the edge counterterms that cancel the
classical divergence is the central subject of this paper.

\subsection{Leading-order terms in the bulk effective action }

The leading-order bulk effective Lagrangian for the Son-Wingate \cite{Son:2005rv} boson $\chi$ is
\bbb
{\cal L}\lrm{SW} \ni c\ll 0 \cc m\uu{{d\over 2}}\cc {\bf X}\uu{1 + {d\over 2}}\ ,
\een{SonWingateLagrangianLO}
where $d$ is the spatial dimension, and $c\ll 0$ is a parameter.
The charge density and Hamiltonian density appear as
\begin{eqnarray}
\r &=& {{\d {\cal L}}\over{\d \dot{\chi}}} = {{\d {\cal L}}\over{\d {\bf X}}} = \Bigl(1+{d\over 2}\Bigr)\, c\ll 0 \cc m\uu{{d\over 2}}\cc{\bf X}\uu{{d\over 2}}\ ,
\nonumber \\
\ \nonumber \\
{\cal H} &=& \dot{\chi}\, \r - {\cal L}  = {d\over 2}\cc {\cal L} + A\ll 0 \r\ .
\end{eqnarray}

In nonrelativistic 
CFT, the parameter $m$ is dimensionless, i.e., inert under rescaling.  It can consistently be set equal to $1$ 
(as some authors \cite{Kravec:2018qnu} choose to do) without breaking conformal invariance, though we leave $m$ indicated 
explicitly throughout the present paper.  In NRCFT with Schr\"odinger symmetry, energy scales
with twice the conformal dimension of spatial momentum, and we will adopt the convention that spatial momentum has scaling
dimension $1$ while energy has scaling dimension $2$, which leaves the spatial coordinates $\vec{x}$ and temporal coordinate $t$
to scale with dimensions $-1$ and $-2$, respectively.

The dynamics of the $\chi$ theory is under perturbative control in 
a homogeneous ground state in infinite volume.  The scattering of goldstone excitations above such a ground state can be computed
reliably in a perturbative expansion, including quantum effects.  To capture quantum effects consistently, subleading terms in the derivative expansion \cite{Son:2005rv} must be included at the appropriate order, while the symmetries of the Schr\"odinger group 
must be preserved by the process of regularization and renormalization.  
The renormalization of the theory is itself under
perturbative control so long as the ultraviolet cutoff $\L$ is set at a scale parametrically higher than the infrared energy scale $E\lrm{IR}$,
and set lower than the bulk strong-coupling scale $E\lrm{UV} \sim \m \propto \langle{\bf X}\rangle \propto \langle\r\rangle\uu{{2\over d}}$:
\bbb
E\lrm{IR} \muchlessthan \L \muchlessthan E\lrm{UV}\ .
\een{DoubleHierarchyForPerturbativeRenormalization}

In the limit $E\lrm{IR} \muchlessthan E\lrm{UV}$, this prescription \rr{DoubleHierarchyForPerturbativeRenormalization}
can be imposed consistently. 
Observables, including quantum effects, can be expressed as a series in $E\lrm{IR} / E\lrm{UV}$, with only a finite number of higher-derivative effective terms or loops in Feynman diagrams contributing at a given order in $E\lrm{IR} / E\lrm{UV}$.  In particular, no higher-derivative term or quantum correction
can contribute to the energy of the homogeneous ground state with constant charge density in infinite volume.

\subsection{Leading-order quantities in infinite volume}

On the grounds of dimensional analysis, the energy density
of the charged homogeneous ground state in infinite volume is
goes as
\bbb
{\cal H}\propto 
 m\uu{-1}\cc \r\uu{{{d+2}\over d}} \propto m\uu{{d\over 2}}\cc \m\uu{1+{d\over 2}}\ .
 \eee
 The proportionality relations follow strictly from dimensional
 analysis, scale invariance, and the existence of a thermodynamic
 limit at finite chemical potential $\mu$ and zero temperature.
 The coefficients of proportionality may depend on the theory
 however, and are expressed in terms of the dimensionless,
 theory-dependent constant $\xi$ (\ref{bertschxi}).
 
 Under the conventions of \cite{Nishida:2010tm}, with two scalar species of fermions in 
dimension $d$ (see the discussion above \rr{bertschxi}),
the energy density of the charged homogeneous ground state of the unitary theory in infinite volume, in terms of the
charge density $\r$ and the Bertsch parameter $\xi$, is
\bbb
{\cal H}\ll{\left [ \cc {{\rm charge~density~\r,~}\atop{\rm interacting~NRCFT}} \cc \right ] }  =  \frac{d}{d+2}\,  \xi\,  \e_{\rm FF}\, \r \ .
\eee
The (free) Fermi energy $\e_{\rm FF}$ in dimension $d$, via the $d$-dimensional (free) Fermi momentum $k_{\rm F}$, is
\bbb
\epsilon_{\rm FF} = \frac{k_{\rm F}^2}{2m}  = \frac{1}{2m}\left(2^{d-1}\pi^{d/2}\G(\frac{d}{2}+1) n \right)^{2/d} \ .
\een{fermiEnergyInDimensiond}
The parameter $\xi$ can, in turn, be expressed in terms of the chemical potential
\bbb
\mu = \frac{d+2}{\r\, d}\, \,  {\cal H}\ll{\left [ \cc {{\rm charge~density~\r,~}\atop{\rm interacting~NRCFT}} \cc \right ] }
\een{chemicalPotentialSW}
as
\bbb
\xi = \frac{\mu}{\e_{\rm FF}}\ .
\eee
In terms of $c_0$, 
\bbb
\xi 
= (2\pi)\uu{-1} \,\, \left( \cc      { \frac{1}{2} {\G\left({d\over 2} + 2 \right)}}  \, c\ll 0 \cc \right)\uu{-{2\over d}}\ ,
\xxn{SonWingateXiDefinitionForTwoSpinlessSpeciesOfFermionInArbitraryDimensionRECAP}
c_0 = {2\over{ \G\left({d\over 2} + 2\right)}}\,\, (2\pi)\uu{-{d\over 2}}\,\, \xi\uu{-{d\over 2}}\ .
\een{C0InTermsOfNishidaSonXiRECAPCopiedCorrected}
In $d = 3$ spatial dimensions, this reduces to
\bbb
\xi \biggl|_{d=3} = 2 \frac{(2/15)^{2/3}}{c_0^{2/3}\pi^{4/3}} \ , \qquad
c_0 \biggl|_{d=3} = \frac{2^{5/2}}{15\pi^2\xi^{3/2}} \ .
\een{3dexpressionsForXi}
as in \cite{Son:2005rv}.

\subsection{Leading-order quantities in a harmonic trapping potential}\label{LeadingOrderRelationsReview}

Let us now introduce the harmonic trap.  
The classical solution in the potential \rr{HarmonicTrappingPotentialDef} is just
\bbb
\chi = \m\, t \ , \llsk\llsk {\bf X} = \m - {{m\o\sqd}\over 2} \vec{x}\sqd\ ,
\een{ClassicalProfileForBoldX}
defined within the radius of nonvanishing charge density,
\bbb
{\bf X}(\vec{x}) > 0\ , \llsk |\vec{x}| < R \equiv {{\sqrt{2\m}}\over{\o\sqrt{m}}}\ .
\een{DefOfDropletRadius}


Before working out the explicit formulae for the energies in the harmonic potential in $d$ dimensions, we can
start by deriving the scalings from dimensional analysis.  The Lagrangian and Hamiltonian densities 
scale as $m\uu{{d\over 2}}\cc \m\uu{1 + {d\over 2}}$, and the size of the droplet goes as $R\sim m\uu{-\hh}\o\uu{-1}\cc \m\uu{\hh}$.
So the total energy at leading order goes as $E\sim m\uu{{d\over 2}}\cc \m\uu{1+{d\over 2}} R\uu d \sim \o\uu{-d} \m\uu{d+1}$.
Then the charge scales as $Q \sim {{dE}\over{d\m}} \sim m\uu{- {d\over 2}}\cc (\m / \o)\uu d$, so 
the chemical potential in terms of the charge is $\m\sim \o Q\uu{{1\over d}}$.


With these general scalings in place, we write the explicit leading-order formulae for the 
total Lagrangian, chemical potential, energy, charge, etc.,~by integrating the leading-order Lagrangian density
\rr{SonWingateLagrangianLO} over the region $|x| < R$.  Using \rr{ClassicalProfileForBoldX}, \rr{DefOfDropletRadius},
and the formula for the area of the unit $(d-1)$-sphere,
\bbb
\area{d-1} 
=   2\, {{\pi\uu{{d/ 2}}}\over{\G({d\over 2})}} \ ,
\een{UnitSphereAreaValueRECAP0}
we have
\bbb
L = c\ll 0\frac{ (2\pi)\uu{{d\over 2}}  \G \left(\frac{d}{2}+2\right)}{\Gamma (d+2)} \left({\m\over\o}\right)\uu{d+1} \o
 = {2\over{\G(d+2)}}\, \xi\uu{-{d\over 2}}\, \big ( \cc {{\m}\over\o} \cc \big )\uu {d+1}\o\ ,
\een{CorrectFormulaForLorentzianLagrangianInTheClassicalGroundStateRECAP}
at leading order, where we have used the identity \rr{C0InTermsOfNishidaSonXiRECAPCopiedCorrected} relating $c\ll 0$
to $\xi$.

Differentiating with respect to $\m$, the leading-order relationship between the chemical potential $\m$ and
charge $Q$ in the isotropic harmonic trap with frequency $\o$, is
\bbb
Q= c_0\frac{ (2\pi) ^{d/2}  \Gamma \left(\frac{d}{2}+2\right)}{\Gamma (d+1)} \left(\frac{\mu }{\omega }\right)^d = {2\over{\G(d+1)}}\, \xi\uu{-{d\over 2}}
\, \big ( \cc {\m\over\o} \cc \big )\uu d\ .
\een{SimplyExpressedInverseRelationshipBetweenMuAndQInTheHarmonicTrapRECAPFirstAppearance}
For what follows, it is convenient to retain the inverse expressions for the chemical potential:
\bbb
\m  = \o\, c\ll 0\uu{-{1\over d}}\cc (2\pi)\uu{-\hh}\cc
\bigg [ \cc {{\G (d+1)}\over{ \G({d\over 2} + 2)}}\cc\bigg ]\uu{{1\over d}}\cc
Q\uu{{1\over d}}
= \o\, \xi\uu{\hh} \,
\bigg [ \cc {{\G(d+1)}\over 2}  \bigg ]\uu{{1\over d}} \cc \cc Q\uu{{1\over d}} \ .
\een{SimplyExpressedRelationshipBetweenMuAndQInTheHarmonicTrapRECAPFirstAppearance}
Since $L$ scales as $\m\uu{d+1}$, we have
\bbb
\m\cc Q = (d+1) \cc L\ , \qquad
H =  d\,  L = {d\over{d+1}}\m\, Q\ , \qquad L = {1\over{d+1}}\m\, Q = {1\over d}\cc H\ .
\eee
Thus, the leading-order expression for operator dimensions in terms of $Q$ and either $c_0$ or
the Bertsch parameter $\xi$ is easily obtained:
\begin{eqnarray}
&&\kern-10mm \D\lrm{{{leading}\atop{order}}}(Q) = \frac{1}{\o}E\lrm{{{leading}\atop{order}}}(Q) = \frac{d}{\o} L\lrm{{{leading}\atop{order}}}(Q)
\nonumber \\
\ \nonumber \\
&& =  c\ll 0\uu{-{1\over d}}\cc {d\over{d+1}}\cc (2\pi)\uu{-\hh}\cc
\bigg [ \cc {{\G (d+1)}\over{ \G({d\over 2} + 2)}}\cc\bigg ]\uu{{1\over d}}\cc
Q\uu{{{d+1}\over d}}
= \xi\uu{\hh} \,  {d\over{d+1}}\cc
\bigg [ \cc {{\G(d+1)}\over 2}  \bigg ]\uu{{1\over d}} \cc \cc Q\uu{{{d+1}\over d}}\ .
\end{eqnarray}\label{LeadingOrderFormulaForOperatorDimensionInTermsOfQAndC0}

\subsection{Next-to-leading order (NLO) terms in the bulk effective action}


Some higher-derivaive effective terms contributing beyond leading order in the expansion in $E\lrm{IR} / E\lrm{UV}$
have been worked out in \cite{Son:2005rv}.  The two terms allowed by diffeomorphism invariance
and conformal invariance, which contribute at NLO, are
\bbb
{\cal L}\lrm{NLO} = {\cal L}\lrm{c1} + {\cal L}\lrm{c2} + 
{\rm (other)} \ .
\een{NLOStructureHere}
Defining
\bbb
{\bf Y} \equiv (\vec{\pp} {\bf X})\sqd\ ,
\xxn{YDef}
 {\bf Z} \equiv  [ \vec{\pp}\sqd\cc A\ll 0 - {1 \over{ d^2\cc m}} (\vec{\pp}\sqd\cc \chi)^2 ]\ ,
\een{ZDef}
we have
\bbb
{\cal L}\lrm{c1} \equiv\cc c_1\, m\uu{\hh(d-2)}\cc {\bf X}\uu{{d\over 2} - 2}\cc{\bf Y}\ ,
\xxn{NLOTerm1}
{\cal L}\lrm{c2} \equiv - c_2\, d\sqd\cc m\uu{\hh(d-2)}\cc {\bf X}\uu{{{d\over 2}} - 1}\cc{\bf Z}\ ,
\een{NLOTerm2}
while the ``other" terms are further suppressed in the large-charge expansion.
For the sake of clarity in the context of the existing literature, it is worth noting that this convention (namely, the assignment 
of $c_n$ coefficients to certain terms in the theory) aligns with that of \cite{Son:2005rv}, among others (albeit in general dimension $d$).  
Other authors \cite{Kravec:2018qnu} make different choices while retaining the $c_n$ notation, so some caution is in order.

Since the higher-order terms come with two additional spatial
derivatives relative to the leading term, they are dressed
to conformality, and so are of relative order $1 / (R\sqd {\bf X}) = O(\o\sqd / \m\sqd)$
compared to the leading term.  In terms of $Q$, using
eqn.~\rr{SimplyExpressedRelationshipBetweenMuAndQInTheHarmonicTrapRECAPFirstAppearance}, they 
are of relative order $O(Q\uu{-{2\over d}})$ compared to the leading term.  As for edge operators, we shall
see in Sec.~\ref{DressedIdentityExample} that the leading
operator has a $\m$-scaling $\m\uu{{2d-1}\over 3}$ after
integration, which is the same size as the integrated subleading bulk operator in $d=2$, and is 
strictly smaller in $d > 2$.  In all, the relative size of corrections from
subleading operators is always of relative order ${{\o\sqd}/{\m\sqd}} = O(Q\uu{-{2\over d}})$.

The authors of \cite{Son:2005rv} justify the restriction to these terms by the symmetries of the theory.  
In addition to the obvious translational, scale, and galilean symmetries, 
\cite{Son:2005rv} demands a conformally invariant and nonrelativistically-``generally-covariant" coupling to background fields,
such as the gauge field and metric. 
In this paper we are going to focus primarily on the first-order
contributions of the $c\ll 1$ term to illustrate the universality
of renormalization of NLO interactions in $d=2$, though in section \rr{EffectOfTheOtherBulkTermInTwoDimensionsAtLeast}
 we briefly consider the first-order contribution of the $c\ll 2$ term as well.

\subsection{Criteria for admissible terms}

At this point, it is sensible to enumerate, at the broadest level, the criteria for allowed higher-derivative terms in the effective action.
This will help to establish structural boundaries for the subsequent organization of allowed operators when we 
eventually address the space of viable counterterms at the droplet edge.  
Let us emphasize that, for the sake of completeness, we are including some criteria here that ultimately do not play such a central
role in the classification of possible terms at the droplet edge.  
In constraining available edge terms, for instance, we need not appeal heavily to diffeomorphism invariance, or conformal invariance beyond rigid scale 
invariance.\footnote{To be sure, in the {\it bulk} theory at NLO we need conformal invariance explicitly to eliminate one otherwise-admissible term.}
These considerations generally enter in very generic ways (like specifying that ${\bf Z}$ and ${\bf Y}$ are conformal primaries, say, or  
appealing to the fact that the dressing rule on the edge requires delta-function support expressed as $\d({\bf X})$, rather than $\d(r - R)$,
which is implicitly a consequence of diffeomorphism invariance).  Even so, these criteria play a role, and it is useful to elucidate the extent of their
influence.

\subsubsection{Gauge invariance}

The internal symmetry $\chi \to \chi + ({\rm const.})$ of the \textsc{sw} theory can be understood
as a particle number symmetry, which enforces mass conservation in a nonrelativistic system.  
This symmetry has a special role, appearing as a commutator in the algebra of
generators of the Schr\"odinger symmetry \cite{Nishida:2007pj}.
Like any internal symmetry of a quantum field theory, one can couple it to a background gauge connection, even though it is a global
symmetry.  When coupled to a background gauge field, the global symmetry can be promoted to a `local gauge symmetry' in the appropriate
sense, so long as the conservation of the original current is exact. To be careful, by local gauge symmetry we do not mean that the action
is invariant under a local symmetry transformation of the dynamical fields alone.  Rather, the system has a local gauge symmetry
in the `spurionic' sense, in which the transformation acts both on the dynamical fields and the background gauge field.

The \textsc{sw} theory can be thought of as an effective theory of the phase variable $\chi$ of the complex fermion field,
\bbb
\psi =e^{-i \chi} \cc \sqrt{\psi\dag\psi}\ .
\eee
In this role, $\chi$ transforms uniquely under a gauge transformation $\a(x,t)$:
\bbb
\chi\to \chi + \a\ .
\eee
To build a covariant Lagrangian for $\chi$ under this local symmetry, we must promote partial derivatives to covariant derivatives,
\bbb
\pp\ll\m\chi\to \gg\ll\m\chi \equiv \pp\ll\m\chi - A\ll\m\ ,
\eee
with the nondynamical background field $A\ll\m$ transforming as
\bbb
A\ll\m\to A\ll\m + \pp\ll\m\a\ .
\eee

\subsubsection{Diffeomorphism invariance}

Other than topological field theories, quantum field theories are never diffeomorphism invariant in the fully dynamical sense.
That is, non-topological QFTs are never invariant under a general diffeomorphism transformation 
acting on the dynamical fields alone.  Rather, in non-topological QFT, diffeomorphism invariance is meant
in the same `spurionic' sense discussed above, in which theories with global symmetries can be 
made covariant under local gauge transformations.  
I.e., the action for the dynamical fields and (non-dynamical) background metric is invariant under a combined transformation of
both, rather than a transformation of the background metric alone.
For relativistic quantum field theories, invariance in this sense under diffeomorphisms connected
to the identity is equivalent to the conservation of the stress tensor.

Nonrelativistic theories are clearly not fully diffeomorphism-invariant, even in this spurionic sense; the formulation of a nonrelativistic
theory intrinsically involves singling out a particular timelike direction.  However, it was shown \cite{Son:2005rv} that an even 
weaker version of the spurionic diffeomorphism symmetry usefully constrains the interactions of the system.  By introducing
a metric $g\ll{ab}$ on vectors pointing in a purely spacelike direction, one can covariantize the system in the canonical
way under time-independent diffeomorphisms $\d x\uu a = \xi\uu a(x)$ of the spatial coordinates $x\uu a$.  By introducing 
additional terms, proportional to $\dot{\xi}\uu a$, in the transformations of the metric and background gauge 
connection $A\ll \m$, one may covariantize the system further under spatial diffeomorphisms depending on time, $\d x\uu a = \xi\uu a(x,t)$.
(For details of the covariantization and useful elements of nonrelativistically-diffeomorphic tensor 
calculus, see \cite{Son:2005rv}.)  It is in this sense that `diffeomorphism invariance' is imposed
as a symmetry of the large-charge effective action for $\chi$ in \cite{Son:2005rv}, which functions as a 
nontrivial constraint on terms in the EFT, beyond the constraints imposed by galilean symmetry and
scale invariance.

\subsubsection{Conformal invariance}

Beyond spatial diffeomorphism invariance, one can also consider reparametrizations of the time coordinate, together
with a time-dependent rescaling of the metric and gauge field,
\bbb
\d t = -t\ ,\llsk\llsk
\d\cc g\ll{ab} = -  \cc g\ll{ab}\ , 
\xxnn
\d A\ll i = 0\ ,\llsk\llsk \d \cc A\ll 0 =  + A\ll 0\ .
\eee
The symmetry algebra of NRCFT contains rigid scale transformations and directly generalizes relativistic conformal symmetry to the nonrelativistic case.  This nonrelativistic conformal group preserves the leading-order action \rr{SonWingateLagrangianLO} of the \textsc{sw} theory \cite{Son:2005rv}, and constrains its possible higher-derivative corrections as well.

\subsection{The dressing rule for bulk operators }\label{BulkDressingRule}

\subsubsection{General comments on dressing rules}

Implicit in the construction of the effective theory is a {\it dressing rule}, controlling which singular local functionals of the fields are allowed to
appear as effective terms in the action.  More precisely, this is a rule specifying which composite or 
composites of dynamical fields can appear as denominators.
A dressing rule of some kind is always a logically necessary ingredient in the construction of any conformally invariant effective Lagrangian.  
Since conformal invariance is never broken directly by the dynamics, there is no external dimensional parameter that can cancel the
conformal dimensions of numerators of higher-derivative terms in the effective action, so the denominators must be some sort of
dynamical field or composite thereof.

Given the central role of the principle of the dressing rule in our subsequent analysis, it is helpful to mention a few contextual examples.
One familiar case is that of the dressing rule in the effective actions for superconformal
theories with vacuum manifolds when the conformal symmetry is spontaneously broken by an expectation value of the vacuum moduli.
In simple cases, such as a vacuum manifold of complex dimension $1$ in ${\cal N} \geq 2$ superconformal symmetry in $D\geq 3$, 
or ${\cal N} \geq 1$ superconformal symmetry in $D\geq 4$, the dressing rule is always uniquely determined.  In such theories,
numerators of vanishing $R$-charge and scaling dimension $\D$ are always 
dressed with the factor 
\be
(\phi\st \phi)\uu{ - {{\D - D}\over{2\cc \D\lrm{\phi}}}}\ ,
\ee 
where $\D\lrm{\phi}$ is the conformal dimension of the modulus $\phi$.   Similar
rules hold in superconformal theories where the vacuum manifold is of the minimal dimension 
allowed by the unbroken and spontaneously broken symmetries of the system, such as
superconformal gauge theories with rank-one gauge 
group \cite{Hellerman:2017veg, Hellerman:2017sur, Hellerman:2018xpi, Hellerman:2020sqj}.

In simple nonsupersymmetric CFT with global symmetries, the dressing rule for the large-charge EFT is also frequently uniquely
determined.  For instance, in the large-charge EFT of the Wilson-Fisher critical $O(2)$ model, the dressing field
is $|\pp\chi|$, where $\chi$ is again the phase of the complex field; the same dressing rule holds for other large-charge
limits in the same universality-class, such as the $\IC\IP(n)$ models at large topological charge \cite{delaFuente:2018qwv}.  
Analogous dressing rules hold for the higher $O(2N)$ models at large Noether charge \cite{Alvarez-Gaume:2016vff}, 
and for the worldsheet CFT of effective 
strings \cite{Polchinski:1991ax,Aharony:2013ipa,Hellerman:2014cba,Hellerman:2016hnf,Sonnenschein:2020jbe}.

The general prescription for a dressing rule in a large-charge EFT is as follows: 
Consider all scalar conformal primaries $\{ {\bf X}\ll i\}$, with conformal dimensions $\D\ll i$, having nonzero
expectation values in the charged ground state.  
To each such operator, assign a $Q$-scaling exponent $\b\ll i$, meaning that the vev
of each ${\bf X}\ll i$ scales as $Q\uu{\b\ll i}$ in the ground state with large charge density $Q$.  
The fields ${\bf X}\ll i$ that maximize the ratio $\b\ll i / \D\ll i$ are the ones that participate 
in the dressing rule.  If there is only one such field that is algebraically independent
in the low-energy Hilbert space, then that field is the unique dressing field.

Ultimately, the rationale for the dressing rule is the principle of naturalness.  The dressing field, raised to the power ${1/\D}$,
gives the scaling of the energy $m\lrm{heavy}$ of the heavy excitations above the large-charge ground state (which 
are integrated out).  The mass formula must be invariant under all symmetries of the system and, at large charge, will be dominated by
the invariant with the largest ratio $\b\ll i / \D\ll i$. 

To be sure, there are known examples in which
the dressing rule is {\it not} unique, including ${\cal N} = 2,\ D=4$ theories with gauge group of rank greater than $1$, 
and the large-charge EFT of nonsupersymmetric theories with Abelian global symmetry of rank greater than $1$ \cite{Monin:2016jmo}.  
In such theories, the dressing field can involve an unknown function of dimensionless ratios of dressing fields.  
Such EFTs, though still having some predictive power, are more weakly constrained relative to theories with a unique dressing rule.
For further general comments on dressing rules, see the related discussions in \cite{Hellerman:2014cba,Hellerman:2016hnf,Hellerman:2017sur} 

\subsubsection{Dressing rule in the Son-Wingate EFT}

In the nonrelativistic case the criterion is the same: Candidate dressing fields are all fields with a vev, 
with the maximum ratio of $Q$-scaling exponent to conformal dimension.
For the Son-Wingate EFT \cite{Son:2005rv}, the field ${\bf X}$ has conformal
dimension $d+2$ and $\m$-scaling exponent $\m\uu 1$, which works out to $Q\uu{{2/ d}}$ in the homogenous
ground state, or $Q\uu{{1/ d}}$ in the harmonic trap.  All other invariants have a lower ratio of $\m$-scaling exponent to
conformal dimension, making ${\bf X}$ the unique dressing field (where it is nonvanishing).
Thus, the dressing rule in the bulk states that the only
singular functionals allowed to appear are singular powers of ${\bf X}$ itself, dressing nonsingular monomials in $\chi$ and
its appropriately covariantized derivatives.

This structure must change at the droplet edge, where ${\bf X}$ vanishes. 
As we will show in Sec.~\ref{DropletEdgeOps}, the operator structure of the Hamiltonian is reorganized on the droplet edge under
a different dressing rule, with $(m\sqd {\bf Y} )\uu{{1/ 3}}$ playing the role of the dressing field for edge operators that ${\bf X}$ plays for bulk operators.

\subsubsection{Bipartite decomposition and $\m$-scaling of bulk operators}

The ${\bf X}$-dressing rule implies a bipartite decomposition for
operators in the EFT:
\bbb
\co = {\bf X}\uu{-p} \co\lrm{undressed}\ ,
\een{XDressingRuleForBulkOperators}
where $\co\lrm{undressed}$ is a polynomial in $\chi$ (and its derivatives), 
plus terms involving couplings to gauge and metric background fields:
\bbb
\co\lrm{undressed} = \sum \prod\ll A[ \pp\ll x\uu {m\ll A} \pp\ll t\uu {n\ll A} \chi]
+ ({\rm background~couplings})
\eee
where $\pp\ll x\uu m$ is short for a linear combination of operators of the form $\pp\ll {x\ll {i\ll 1}} \pp\ll{x\ll{i\ll 2}}\cdots \pp\ll{x\ll{i\ll m}}$.  The dimension
of the operator above is
\bbb
\D\lrm{und} \equiv \D(\co\lrm{undressed}) =  \sum\ll A (m\ll A + 2 n\ll A)\ ,
\eee
and the dimension of the corresponding dressed operator is
\bbb
\D(\co) = \D\lrm{und} -2p\ .
\eee
An operator that can appear in the conformal EFT lagrangian density
must be gauge invariant and must also be a conformal scalar primary
of weight $d+2$.  This criterion determines the exponent $p$ of
the ${\bf X}$-dressing:
\bbb
p = \hh\D\lrm{und} - (1 + {d\over 2})\ .
\eee

\subsection{Equations of motion}\label{EOMReview}

Varying the leading order action with respect to $\chi$, we find
\bbb
0 = - \pp\ll t \big ( \cc {\bf X}\uu{{d\over 2}} \cc \big ) + {1\over m}
\cc \pp\ll i \cc \big [ \cc (\pp\uu i \chi) \cc {\bf X}\uu{{d\over 2}} \cc \big ]\ .
\eee
Note that the leading-order EOM allows us to eliminate temporal derivatives of ${\bf X}$ in favor of terms with just spatial derivatives.  
This is useful when one wants to count independent operators in the action beyond leading order.
Linearizing the EOM in a trivial background, $\chi = \m t + \hat{\chi}$, we have
\bbb
\ddot{\chh} =  {{2\m}\over{{d\cc m}}}  (\gg\sqd\chh) + O(\hat{\chi}\sqd)\ ,
\eee
so the leading-order dispersion relation is
\bbb
\o\sqd = c\ll s\sqd \cc p\sqd\ ,
\llsk\llsk
c\ll s\sqd = {{2\m}\over{d\cc m}}\ .
\een{DispersionRelationInfiniteVolumeTrivlalBackground}

\subsubsection{NLO corrections to the dispersion relation}

Note that the dispersion relation is not fixed by conformal symmetry.
Higher-derivative corrections at NLO and beyond that are consistent
with conformal symmetry affect the dispersion relation by
terms proportional to the coefficients $c\ll {1,2,\cdots}$
For instance, the leading-order dispersion relation \rr{DispersionRelationInfiniteVolumeTrivlalBackground}
is corrected \cite{Son:2005rv, Favrod:2018xov} as
\bbb
\o(p)= c\ll s \cc |p| + \big ( \d\cc  \o(p) \big ) \lrm{NLO} \ , 
\xxx
\big ( \d\cc  \o(p) \big ) \lrm{NLO}   \equiv
- {{(2\pi)\uu{d/2}\cc \Gamma({d\over 2})}\over{\sqrt{2\cc d}}}\,  \xi\uu{{d\over 2}}  \left( c\ll 1 + {d\over 2} \cc c\ll 2\right) \, {{p\uu 3}\over{m\uu{{3/2}}\cc \m\uu{{1/2}}}}\ ,
\een{GeneralDimensionNLOContributionToDispersionRelationSUMMARIZED}
at leading order in $c\ll{1,2}$.  Note that our
conventions for the normalizations of $c\ll{0,1,2}$ and $\xi$ agree with those
of \cite{Son:2005rv, Favrod:2018xov}\footnote{We, like \cite{Son:2005rv}, set $\hbar\to 1$, while
\cite{Favrod:2018xov} leave $\hbar$ dimensionful.}, and we have also written the quantity $d\ll 0\sqd$,
defined in \cite{Favrod:2018xov}, in terms of $\xi$ rather than $c\ll 0$ (see eq. \rr{ValueOfLittleDNotSquaredFromFavrodEtAl} in the Appendix).

\subsubsection{First-excited spectrum in the harmonic potential}

Now consider the EOM in the harmonic potential \rr{HarmonicTrappingPotentialDef}, which describes nonnegatively
charged local operators in radial quantization in NRCFT \cite{Nishida:2007pj}.
In the harmonic potential $A_0 = \frac{1}{2}\o^2 {\vec{x}}^2$, the linearized EOM reads:
\bbb
0 = \ddot{\hat\chi} - \frac{2}{m d}(\m - A_0) \nabla^2 \hat\chi + \frac{1}{m}\o^2 \vec{x}\cdot \vec{\pp}\hat\chi 
\een{HarmonicPotentialDispersionRelation}
Using the dispersion relation \rr{HarmonicPotentialDispersionRelation}
 in the harmonic potential, the linearized excitation spectrum about the ground state was computed in \cite{Kravec:2018qnu}. 
 The result is a first-excited spectrum labelled by integers $n ,\ell \geq 0$, with energies
\bbb
\e(n,\ell) \equiv   \o\cc \left( \cc 
{4\over d} \cc n\sqd + 4n + {4\over d} \ell n - {4\over d} n + \ell
\cc \right) \uu\hh\ .
\een{PhononEnergySpectrum}
For $\ell = 1$ and $n=0$, this has energy $+\o$, and for $\ell = 0$ and $n=1$ it has energy $+2\o$,
so these are precisely the conformal
raising operators corresponding to $\vec{\pp}\ll x$ and $\pp\ll t$, respectively, under the NRCFT state-operator correspondence
reviewed in sec. \ref{StateOpCorrNRCFT}.

This spectrum gives the leading-order spectrum of primaries with dimension $O(1)$ above the ground state.  
The sum over these linearized frequencies, times a factor of ${1\over{2\o}}$, also
gives an $O(Q\uu 0)$ contribution to the ground state energy via the Coleman-Weinberg formula.  
Sometimes, as in $d=2$, there are no bulk or boundary counterterms of order $Q\uu 0$, 
and this contribution is universal after renormalization of bulk and boundary divergences.  
The details of the renormalization of loop divergences is beyond the level of detail we will attempt in the present 
article, and we defer it to future work \cite{toAppearSHIS}.  However, we will give a crude bound on the size of quantum
effects of edge operators in sec. \rr{CrudeUpperBoundQuantumScalings}, 
and we will briefly discuss the implications of that bound in sec.~\ref{conclusions}.  

\subsection{Comments on bulk terms beyond NLO}
 
We can use the ${\bf X}$-dressing rule \rr{XDressingRuleForBulkOperators} for bulk operators to estimate the effect of
higher-derivative terms in the action on observables at a given
length, time, momentum, or energy scale.  For a process characterized by
a momentum scale $p$, a NLO term makes a subleading contribution suppressed by $p\sqd / \m$ for every
two additional spatial derivatives in the term,\footnote{Note again that one loses no generality by restricting to spatial derivatives; 
we can always eliminate time derivatives by the leading-order EOM.} which come along with an
additional ${\bf X}\uu{-1} \sim \m\uu{-1}$.   For a process characterized by a 
length scale $L$, we have a factor of $(L\sqrt{m\m})\uu{-1}$ for
each additional derivative.  So, for instance, an operator in the Lagrangian of the form
\bbb
{\cal L} \ni \co \equiv \k\cc (\pp\ll x {\bf X})\uu{2k} / {\bf X}\uu{3k-{d\over 2} - 1}
\eee
contributes to correlators as
\bbb
\langle \cdots \rangle\ll{O(\k\uu 1)} = 
\langle \cdots \rangle\ll{O(\k\uu 0)} \times (L\sqd m \m)\uu{-k}\ ,
\eee
for a correlation function $\langle \cdots \rangle$ characterized by the distance
scale $L$.  We denote this property by saying that the
operator $\co \equiv (\pp\ll x {\bf X})\uu{2k} / {\bf X}\uu{3k-{d\over 2} - 1}$
has $\m$-scaling $\m\uu{1 + {d\over 2}-k}$, to be understood as an abstract property
of a term in the EFT action.  Implicitly, the dimension will be compensated 
by powers of the infrared scale, which will be taken to lie parametrically 
below the UV scale set by $\m$.

To further simplify the discussion, 
we can also leave out the factor $\m\uu{1+{d\over 2}}$ of the $\m$-scaling 
of the leading order term ${\bf X}\uu{1+{d\over 2}}$, and speak of
terms as having a {\it relative} $\m$-scaling against leading-order quantities
($\m\uu{-k}$ in the example at hand).  
The relative $\m$-scaling
gives us a simple way to further distinguish between NLO terms
in the action in order of importance.  
Thus, both terms \rr{YDef},\rr{ZDef} in the NLO action \rr{NLOStructureHere} have relative $\m$-scaling
$\m\uu{-1}$, and should contribute to any (renormalized) observable with an additional power of $p\sqd / \m$ or $1 / (\m L\sqd)$ for each insertion
of a power of the NLO term.  

This assignment of $\m$-scalings to operators treats the IR scale
as being independent of $\m$.  This is useful for infinite-volume computations
and the local study of correlation functions, so we might refer to it as
the ``infinite volume" $\m$-scaling if we want to distinguish such $\m$-scalings from cases 
in which the natural infrared scale also has a nontrivial scaling 
with $\m$ (as with global observables in the harmonic potential).
Also note that the $\m$-scalings we have discussed here apply only to renormalized observables.
For bare observables, the $\m$-scaling 
is the same but the momentum $p$ may be replaced with a power of
the momentum cutoff or inverse distance cutoff.   In EFT, this
is normally relevant only for quantum processes, but for backgrounds with
classical singularities, such as the edge of the density distribution in a
harmonic potential, a cutoff may appear in the contributions of
NLO terms to classical processes as well.  It is this sort of UV-divergent contribution that
is of particular importance for our present goals.

\subsubsection{$\m$-scalings of higher-derivative terms in the harmonic potential}

From the discussion above, it is clear that the bulk expansion is an expansion
in $1 / (m\m L\sqd)$, where $L$ is the characteristic
distance scale of a process.  We now use this general analysis
to estimate the effect of higher-derivative terms 
on the ground state energy in the harmonic potential.  In this background,
the only available distance scale $L$ is the size of the
droplet $R=  \sqrt{2\m} /(\o\sqrt{m})$ 
on which the charge density is supported \rr{DefOfDropletRadius}.  So,
in the harmonic potential, it is useful to assign a given NLO term
a $\m$-scaling that includes the $\m$-scaling of the IR distance scale
$R$ as well.  Per the note of caution above, 
this is a different $\m$-scaling than the ``infinite volume"
$\m$-scaling we have already discussed, which applies to observables
in which the relevant infrared scale is $\m$-{\it independent}.

In the infinite-volume $\m$-scaling, the leading-order action
scales as $\m\uu{1 + {d\over 2}}$.  In the harmonic-potential $\m$-scaling,
one also has to include the volume of the spatial integration region,
which goes as $L\uu d = R\uu d \propto \m\uu{+{d\over 2}}$, so
the harmonic-potential $\m$-scaling of the leading-order action is
$\m\uu{d+1}$.  

Next, we can consider subleading terms in the action.  For every two additional spatial derivatives in a NLO term in the action, there is an additional
factor of $1 / (m\m L\sqd) \propto \o\sqd / \m\sqd$ in its contribution to an observable, relative to the leading-order contribution.  So the $\m$-scaling
of an integrated NLO term in the harmonic potential is two powers of
$\m$ less for every additional two spatial derivatives in the numerator.

Denoting the bulk harmonic-potential 
$\m$-scaling of an (unintegrated) operator by $\a(\co)$, we have
\bbb
\a({\bf X}) = 1 \ ,
\llsk
\a({\bf Y}) = 1\ ,
\llsk
\a({\bf Z}) = 0\ .
\eee
Note that these $\m$-scalings can be read off directly from the behavior of the classical solution as we take $\m$ large while
keeping the location $x$ of the operator fixed inside the droplet at $|x| < R$.

For conformally dressed operators, we have
\bbb
\a({\bf X}\uu{ {d\over 2} + 1}) =  {d\over 2} + 1\ ,
\llsk
\a({\bf Y} \cc {\bf X}\uu{{d\over 2} -2}) =  {d\over 2} - 1\ ,
\llsk
\a({\bf Z} \cc {\bf X}\uu{{d\over 2} -1}) =  {d\over 2} - 1\ .
\een{BulkMuScalingsHarmonicPotential}
For {\it integrated} conformally dressed operators, we simply
include the additional factor of the size $R\uu d\propto \m\uu{{d\over 2}}$ of
the droplet, to get
\bbb
\a\lrm{integrated}({\bf X}\uu{ {d\over 2} + 1}) =  d + 1\ ,
\xxnn
\a\lrm{integrated}({\bf Y} \cc {\bf X}\uu{{d\over 2} -2}) =  d- 1\ ,
\xxnn
\a\lrm{integrated}({\bf Z} \cc {\bf X}\uu{{d\over 2} -1}) =  d - 1\ .
\een{BulkMuScalingsHarmonicPotentialIntegratedOverDroplet}
 In general, every pair of spatial derivatives in a conformal
 bulk term must come with an additional
$(\o\sqd / \m\sqd)$ in the harmonic potential,
so for conformal terms with no time derivatives we have
\bbb
\a\lrm{integrated}(\co) = d+1 - \#(\pp\ll x)\ ,
\xxnn
\int d\uu d x \cc \co \sim \big [ (\m / \o)\uu{d+1 -   \#(\pp\ll x)}\big ] \times \o\ .
\een{IntegratedBulkTermMuScaling}
In $d=2$, we see that for terms without spatial derivatives the only
potential contributions larger than order $\m\uu 0\leftrightarrow 
Q\uu 0$ come from terms with two derivatives, which have already been examined
in \cite{Son:2005rv}, and written here in \rr{NLOStructureHere},\rr{YDef},\rr{ZDef}.
For $d=3$, there are four-derivative terms that can contribute classically
at order $Q\uu 0$, the same 
order as the first quantum correction.  The same is always true
in any odd spatial dimension: there are always $(d+1)$-derivative
terms that contribute at the same order as the one-loop vacuum energy,
order $Q\uu 0$. 

In even spatial dimensions there are no bulk terms contributing at order $Q\uu 0$ to the renormalized vacuum energy; in some cases,
such as $d=2$, we will see further that there are no edge terms that ever contribute at order $Q\uu 0$
either, indicating that the order $Q\uu 0$ term is universal and calculable
in $d=2$.

\section{Structure of local operators at the droplet edge}\label{DropletEdgeOps}

\subsection{General comments}

As noted earlier, the phenomenon of localized singularities in classical solutions describing large-quantum-number states
is ubiquitous in the subject of large-quantum-number, or LQN, expansions.
The primary difference between the droplet edge and a fixed-position boundary is that the position of a true boundary is just that: 
it is fixed.  Indeed, the dynamical fluctuations of the droplet edge are incorporated in the fluctuations of the $\chi$ field, within the regime of validity of the large-charge EFT.

In the case of the open effective string with Neumann boundary conditions,
it was understood \cite{Hellerman:2016hnf} that the proper way to interpret these singularities is not as a signal that the EFT breaks down altogether,
but that the organization of operators changes, wherein a different dressing rule applies for the denominators of operators in the singular
region, relative to that in the bulk.\footnote{For a related discussion in the case of fold singularities in effective strings, see \cite{Sonnenschein:2020jbe}.} 
Qualitatively, this can be understood in terms of the parametric scaling of energies of heavy excitations at large charge, which are integrated out.  
In the singular region, these heavy modes are parametrically lighter than the corresponding heavy modes in the bulk,
but they still have energies scaling with a positive power of the charge or chemical potential.  For the CFT to break down,
the energies of the heavy modes would have to go to zero, but they do not.  In the singular region, the dressing field is simply
proportional to the local expression describing the parametric energy scale of the heavy modes.  In practice, this means
the dressing field is always a power of the EFT operator with the largest $Q$-scaling (or $\m$-scaling) per conformal dimension, both
in the bulk and in the singular region.\footnote{When such an operator is unique, 
as in \cite{Hellerman:2015nra, Monin:2016jmo, Alvarez-Gaume:2016vff, Hellerman:2017efx, 
Polchinski:1991ax, Hellerman:2013kba, Aharony:2009gg, Aharony:2010cx, 
Hellerman:2016hnf, Aharony:2010db, Aharony:2011ga, Aharony:2011gb, Hellerman:2017veg, 
Hellerman:2017sur, Hellerman:2018xpi, Hellerman:2020sqj} , there is a unique choice of dressing field for the large-charge EFT. The dressing operator is not always unique, however,
as in the case of a generic CFT with $U(1)\sqd$ symmetry \cite{Monin:2016jmo}.  In these cases the EFT still has predictive power,
but somewhat less so absent additional information.}


Despite the dynamical nature of the droplet edge, as opposed to a fixed boundary,
we can still treat the droplet edge in EFT because
the fluctuations in its position admit an energy 
cost that can be estimated according to the phonon energy spectrum \rr{PhononEnergySpectrum}. 
These energies are $O(\o\uu 1\m\uu 0)$  for phonons whose angular momenta $\ell$ and radial
wavenumbers $n$ are $O(1)$.  The low-energy Hilbert space 
can accommodate only some fixed number of such excitations
(depending on where one places the cutoff $\L$), so the fluctuations
of the coordinate position of the boundary are actually rather small.
Thus, we are safe treating the position of the boundary as a 
semiclassically fixed entity, so long as we incorporate the effects of
the fluctuations of its position systematically in perturbation theory,
while maintaining all the symmetries of the system.

The formal method for doing this generalizes the way we treat boundary
or defect operators.  For a boundary operator, we write
$\Delta {\cal L} = \d(\s\ll 1 = \s\ll 1\upp 0) \cc \co(\s)$, if $\s$ are 
the coordinates and the boundary is normal to the $\s\ll 1$ direction.
For a droplet-edge operator, we accommodate
the fluctuations of the droplet edge by letting
the argument of the $\d$-function be the dynamical field ${\bf X}$
rather than a coordinate:
 \bbb
 \Delta {\cal L}\lrm{edge} = \d({\bf X}) \cc \co(x)\ .
 \eee
The large-$Q$ expansion is of course only an asymptotic
 expansion, but in this expansion we can understand $\d({\bf X})$
 concretely as a boundary operator at the classical droplet
 edge $R = {{\sqrt{2\m}}/({\o\sqrt{m}})}$, plus a series
 of higher-derivative boundary operators that are parametrically
 suppressed at large $\m$.
 While expressions like $\d({\bf X})$ appear to be singular,
 we proceed with the understanding that the 
 EFT is regularized and renormalized, and all local operators are defined
 at an energy scale $\L$ that is parametrically below the scale set by the
 chemical potential.  We give a concrete expansion of $\d({\bf X})$ in
 fluctuations in sec.~\ref{PhononExpansionOfDeltaFunctionOfBoldX}. 
 
\subsection{The ${\bf Y}$-dressing rule for edge operators}\label{YDressingRule}

In sec.~\ref{BulkDressingRule} we discussed the bulk dressing rule.  Namely, the only 
field allowed to appear in the (bulk) effective action raised to negative or fractional powers is the 
composite ${\bf X}$ itself.  The appearance of such functions of ${\bf X}$ are innocuous so long as one does not try to 
use the effective theory in a regime where ${\bf X}$ is small compared to the relevant infrared energy scale.  
Of course, this is precisely the problem for describing the complete system with finite droplet extent, since 
${\bf X}$ vanishes at the edge of the trapped droplet.  Thus, we need to adopt an applicable dressing rule at the edge.
The dressing operator for {\it near-edge} operators should be the operator with the lowest $\m$-scaling per unit conformal dimension
that is nonvanishing at the edge.  This is the operator ${\bf Y}$, defined in eq.~\rr{YDef}:
\bbb
{\bf Y} \equiv  (\vec{\pp} {\bf X})\sqd\ .
\een{YDefRecap}
The ${\bf Y}$-dressing rule at the edge thus
allows terms such as $\co / {\bf Y}\uu p$, where $\co$ is
a polynomial in dynamical and background fields and their derivatives, but not terms such as $\co / {\bf Z}\uu q$, 
with ${\bf Z}$ defined in eqn.~\rr{ZDef} above.

To understand why the former $(\co / {\bf Y}\uu p)$ is allowed for a nonsingular
droplet edge but not the latter $(\co / {\bf Z}\uu q)$, the reasoning is similar
to the reasoning for dressing rules in other cases,
including the ${\bf X}$-dressing rule for bulk operators.
As with the bulk theory, the principle can be understood as a consequence of
naturalness:  The denominators of allowed terms should be powers
of the energies ${\cal E}$ of unknown integrated-out excitations, which should be presumed to have the most 
generic possible energy formula allowed by the symmetries.  
The formula for ${\cal E} = {\cal E}({\bf X}, {\bf Y}, {\bf Z}, \cdots)$
can depend on the local fields in a general way, consistent
with conformal invariance, which forces ${\cal E}$
to be a scalar primary of weight $2$, but with no other constraints.
At a point in the bulk, where ${\bf X}$ is nonvanishing,
a generic energy formula ${\cal E} = {\cal E}({\bf X}, {\bf Y}, {\bf Z}, \cdots)$ will be dominated by the term of largest possible
$\m$-scaling per unit conformal dimension, which is ${\bf X}$.
Only when ${\bf X}$ is vanishing should the formula 
for ${\cal E}$ be dominated by the invariant with
the {\it next} largest $\m$-scaling per unit conformal dimension, i.e.,~${\bf Y}$.

As we shall see, the ${\bf Y}$-dressing rule at the edge
has consequences for the $\m$-scaling
laws associated with edge operators.  Namely,
operators such as ${\bf Y}\uu p {\bf Z}\uu q$
are not allowed at the edge unless $q$ is a nonnegative integer.
This can be understood easily on physical grounds:
Consider the ground state of the system in a constant
electric field ${\cal E}$ in the $x\ll 1$-direction.  For such a
system the ground state solution has ${\bf X} = {\cal E}(x\ll 1 - x\ll 1\upp 0)$.  The physics of the infinite, flat droplet edge at
$x\ll 1 = x\ll 1\upp 0$ is locally identical to the
physics of the circular edge in the harmonic potential,
but the ${\bf Z}$-invariant vanishes identically everywhere
in the ground state of the constant electric field.
The properly chosen dressing rule must give
an organization of operators that makes sense in the neighborhood
of {\it any} edge that is locally of a generic type with well-defined
unit normal vector.  An organization of operators dressed with singular powers of
${\bf Z}$ is ill-defined at the edge of the charge
distribution in a constant electric field, so we conclude that 
operators dressed with fractional or negative powers of ${\bf Z}$
are not allowed.  Similar considerations forbid any dressing
field other than ${\bf Y}$ for a droplet edge of nondegenerate
type with well-defined normal vector at the edge.\footnote{Of course, one can always consider more general 
singular configurations in which the background electric
potential is organized so that the edge has a cusp or corner, or where an open set of the 
locus ${\bf X} = 0$ has a double zero, or otherwise where there exists an ill-defined normal vector at the edge.  
In such geometries, the ${\bf Y}$ field vanishes at the edge and the organization of operators would
be different again.  In the most generic such situations we would
expect a ${\bf Z}$-dressing rule to apply.  However interesting to contemplate, we do not consider
such situations in the present article.}

\subsection{The tripartite structure of droplet-edge operators}\label{CoarseAnalysisOfTermsWithYDressingRuleAtDropletEdgePRECAPSec}

The discussion above thus gives a general recipe for constructing edge operators:
\bi
\item{Start with a nonsingular bulk operator with no undifferentiated
${\bf X}$ fields.  For lack of a better name, let us say that this plays the role of an undressed {\it numerator.}}
\item{Append a factor of $\d({\bf X})$, so that the term is supported
only at the edge of the droplet.}
\item{Further append powers of ${\bf Y}$ to make the operator
a conformal primary of marginal conformal weight.}
\ei
That is, edge operators can be thought of as being
generated by terms with a simple tripartite structure:
\bbb
{\cal O}\lrm{edge} = \d({\bf X}) \cc {\bf Y}\uu {-p} \cc {\cal O}\lrm{numerator}\ .
\eee

Let us now discuss each sector of this structure.  First, ${\cal O}\lrm{numerator}$ is 
an arbitrary monomial in $\chi$ and its derivatives, modulo positive powers of ${\bf X}$ and terms that vanish
by the EOM.  In fact, the properties of ${\cal O}\lrm{numerator}$ are just inherited from the analogous requirements 
pertaining to the operator dressing rule in the bulk theory (see Sec.~\ref{BulkDressingRule}), 
so we need not revisit that discussion in full detail.  The other two parts of the dressing structure 
are unique to the edge theory, and they deserve further commentary.

\subsubsection{The $\d({\bf X})$ factor}


It is convenient to consider the $\d$-function as an explicit part of the dressing rule, since
all edge terms must appear with precisely one power of $\d({\bf X})$.
Formally, the operator $\d({\bf X})$ is a conformal
primary of {\it nonrelativistic} scaling dimension $-2$, where the spatial
partial derivative is counted with weight $1$ and the time derivative
is counted with weight $2$.  The $\d$-function of an object always carries
the negative of the conformal dimension of that object, in any
kind of semiclassical expansion.  Of course, going to higher-order in
insertions of the edge-term itself will force us to regularize the $\d$-function,
but at first order in the insertion of the edge term the regularization is
irrelevant.

The $\d({\bf X})$ factor is more intuitive if one uses the inverse chain rule
for $\d$-functions $\d(f(r)) = {1\over{|f\pr(r\ll 0)|}}\cc \d(r - r\ll 0)$ to
rewrite it as a $\d$-function in $X$-space, multiplied by an operator-valued
measure factor.  
That is, the $\d$-function of an operator $\hat{\co}$ can be understood as
\bbb
\d(\hat{\co}) = \d(r - \hat{r}\ll 0\ups{\co}) \, |\vec{\pp}\ll x \hat{\co}|\uu{-1}\ ,
\een{OperatorDeltaFunction}
where $r$ is simply a coordinate label for local operators, and
$\hat{r}\ll 0\ups{\co} \equiv \hat{r}\ll 0\ups{\co}(\th,t)$ is an operator 
denoting the radial position of the vanishing locus
of $\hat{\co}$ at a particular angular direction $\th$ and time $t$.
For all purposes in the present paper, we will be considering
vacuum expectation values of the edge operators. In the
classical approximation in the vacuum, $\hat{r}\ll 0\ups{\co}(\th,t)$
is a $c$-number, independent of $\th$ and $t$, equal
to the classical radius of the droplet:
\bbb
\hat{r}\ll 0\ups{\co}(\th,t) \simeq R = {{\sqrt{2\m}}\over{\o\cc\sqrt{m}}}\ .
\eee
Thus, in this approximation, we retrieve via \rr{YDef}:
\begin{eqnarray}
\d({\bf X}) &\simeq & \d({\bf X}\lrm{classical}) = |\vec{\pp} {\bf X}\lrm{classical}|\uu{-1}\cc \d(|x| - R)
\nonumber\\ \ \nonumber\\
& =& {\bf Y}\lrm{classical}\uu{-\hh}\cc \d(|x| - R) 
\nonumber\\ \ \nonumber\\
&=& {1\over{m\o\sqd R}} \d(|x| - R) \ .
\label{JacobianFormulaClassicalApprox}
\end{eqnarray}
In section \ref{PhononExpansionOfDeltaFunctionOfBoldX}, we will see how to go beyond this approximation 
to incorporate quantum fluctuations as operator contributions to $\d({\bf X})$.

\subsubsection{The ${\bf Y}$-dressing}

Finally, as discussed above, naturalness (among other considerations) dictates
that droplet-edge terms be dressed to conformality with powers of ${\bf Y}$,
which itself has conformal dimension 6.  
For an edge operator ${\cal O}\lrm{edge}$ appearing in the 
conformal EFT Lagrangian density, with the ${\bf Y}$-factor appearing as
${\bf Y}^{-p}$, conformal invariance of the perturbing Hamiltonian will
thus force $p=-{1\over 6}(d + 4 -  \D\lrm{und})$, where $\D\lrm{und}$ is the dimension of the
undressed {\it numerator} factor $ {\cal O}\lrm{numerator}$.
This exponent is chosen so that the total (momentum) scaling 
dimension of $\co\lrm{edge}$ is equal to $d+2$, with the scaling dimension of $\d({\bf X})$
counted as $-2$.
So, altogether, the droplet-edge terms are of the form
\bbb
{\cal O}\lrm{edge} = \d({\bf X}) \, (m\sqd {\bf Y})\uu{{1\over 6}(d + 4 -  \D\lrm{und})}\, {\cal O}\lrm{numerator}\ .
\een{EdgeOperatorGeneralForm}
In the following sections we will calculate the power law with
which an edge operator of the form \rr{EdgeOperatorGeneralForm} contributes.
I.e.,~we want to characterize the exponent $\g$ such that $\langle \int\cc d\uu d x\cc
\co\lrm{edge} \rangle \lesssim \m\uu\g$ at large charge.

\subsection{Example: A simple droplet-edge operator and its $\m$-scaling}\label{DressedIdentityExample}

The simplest droplet-edge operator is the case where the numerator is simply the identity.  To make the operator conformally
invariant, we need to dress it with ${\bf Y}$ so that the total weight is marginal:
\bbb
{\cal O}\lrm{edge} =  \d({\bf X})\, {\bf Y}\uu{{{d+4}\over 6}}\ .
\een{DressedIdentityEdgeOp}
At the classical level, we can simply replace ${\bf X}$ with its classical profile.  
Per the discussion above, translating the $\d$-function of ${\bf X}$ into a $\d$-function of position in the radial
direction, we obtain $\d({\bf X}) = |{\bf Y}|\uu{-\hh}\cc \d(|x| - R)$. 
The classical ground-state value of ${\bf Y}$ evaluated at $r = R$ is
\bbb
{\bf Y}\lrm{classical} \cc \big |\ll{|x| = R} = m\sqd \o\uu 4 R\sqd = 2\,m\,\o\sqd\m\ .
\een{BoldYInvariantClassicalAtDropletEdgeValue}
The the $\m$-scaling of ${\bf Y}$ is $\sim \m\uu{+1}$, and the $\m$-scaling of $\d({\bf X})$ is ${\bf Y}\uu{-\hh}\sim \m\uu{-\hh}$, 
so, in total, the $\m$-scaling of ${\cal O}\lrm{edge}$ is $\m\uu{{{d+1}\over 6}}$.
Since the size of the integration region is $R\uu{d-1} \propto \m\uu{{d-1}\over 2}$, 
the $\m$-scaling of the {\it integrated} operator is $\m\uu{{{2d-1}\over 3}} $.

\subsection{Expansion in phonons}\label{PhononExpansionOfDeltaFunctionOfBoldX}

To go beyond the classical level, we need to understand the meaning of the expression $\d({\bf X})$ as interpreted in EFT.
Nonperturbatively, there may or may not be some ambiguity in the definition of $\d(\co)$, but it can
always be given a meaningful definition in an effective theory with a specified
regularization procedure.  For instance, if $\co$ vanishes at more than one value 
of $r$ at a given time and angular direction, $\hat{r}\ll 0(\O,t)$ can be defined to be the 
largest value of $r$ at which $\hat{\co}(r,\O,t) = 0$, or some variation on that rule.  
In semiclassical perturbation theory around a classical ground state for
which $\hat{\co}$ vanishes exactly once, at a distance $r = r\ll {0,~{\rm classical}}\ups{\co}$ 
from the origin, the operator $\hat{r}\ll 0\ups{\co}$ can be represented
as 
\bbb
\hat{r}\ll 0\ups{\co} = r\ll {0,~{\rm classical}}\ups{\co} + \widehat{\d r}\ll 0\ups{\co}\ ,
\eee 
where $\widehat{\d r}\ll 0\ups{\co}$ has a vanishing expectation
value in the vacuum, and its fluctuations are suppressed in the semiclassical expansion.

This situation describes exactly the case $\hat{\co} = {\bf X}$ in the harmonic
trap, where the classical vanishing radius $r\ll 0\ups{{\bf X}} = \langle \hat{r}\ll 0\ups{{\bf X}} \rangle$ 
is just the droplet radius $R = {{\sqrt{2\m}}/({\o\cc\sqrt{m}})}$, and the quantum fluctuations of $\hat{r}\ll 0\ups{{\bf X}}$
in the ground state are suppressed by powers of the total 
fermion charge $Q$.
In perturbation theory about the ground state of the harmonic trap
at large $Q$, the fluctuations around the classical configuration
of ${\bf X}$ are small, so the fluctuations $ \widehat{\d r}\ll 0\ups{{\bf X}}$ of $\hat{r}\ll 0\ups{{\bf X}}$ about
the vev $r\ll 0\ups{{\bf X}} = \langle \hat{r}\ll 0\ups{{\bf X}} \rangle = R$ are also small.  We can use the
suppression of fluctuations $\widehat{\d r}\ll 0\ups{{\bf X}}$ to 
represent the droplet-edge operators as a series of operators
supported {\it exactly} at the classical vanishing radius.

To make this concrete, we can break up the operator $\widehat{{\bf X}}$ into vev and fluctuations,
\bbb
\widehat{{\bf X}} = {\bf X}\lrm{cl} + \widehat{\d{\bf X}}\ , 
\eee
where $ {\bf X}\lrm{cl}$ is the classical vev of ${\bf X}$ at some given
chemical potential in the harmonic trap.  We can then write the
$\d$-function as
\begin{eqnarray}
\d({\bf X}) &=& \d({\bf X}\lrm{cl} +  \widehat{\d{\bf X}}) 
\nonumber\\
&=& \d({\bf X}\lrm{cl}) + \d\pr({\bf X}\lrm{cl})\cc \widehat{\d{\bf X}}
+ \hh \cc \d\prpr({\bf X}\lrm{cl})\cc  (\widehat{\d{\bf X}})\sqd
+ O\bigg ( \cc(\widehat{\d{\bf X}})\uu 3 \cc \d\prprpr({\bf X}\lrm{cl})\cc \bigg )\ .
\end{eqnarray}
First, the zero-fluctuation term is simply
\bbb
\d({\bf X}\lrm{cl}) = |\pp\ll r {\bf X}\lrm{cl}|\uu{-1} \cc \d(|x| - R)\ .
\eee
Now, to understand the localized distribution appearing in the one- and 
two-fluctuation terms, we can use various $\d$-function identities and the form of the classical solution
\bbb
{\bf X}\lrm{cl}(r) = \langle  {\bf X}(r) \rangle 
= \m - {{m\o\sqd}\over 2} \cc r\sqd\ ,
\eee
to derive the expansion of the $\d({\bf X})$ operator in fluctuations:
\def\STUFFDEFOUT{
\bbb
f(r) \cc \d\pr(r - r\ll 0) = f(r\ll 0)\d\pr(r - r\ll 0) - f\pr(r\ll 0) \cc \d(r - r\ll 0)\ ,
\een{DeltaFunctionPrimeProductIdentity}
we get
\bbb
\d\pr({\bf X}\lrm{cl}) =- {1\over{m\sqd\o\uu 4 \cc R\sqd }} \cc \d\pr(|x| - R)
- {1\over{m\sqd\o\uu 4\cc R\uu 3}}\cc \d(r-R)
\eee
and then from there we can get
\bbb
\d\prpr({\bf X}\lrm{cl}) = {1\over{\pp\ll r \cc {\bf X}}} \cc {d\over{dr}}
\cc\big [ \cc  \d\pr({\bf X}\lrm{cl}) \cc \big ]
\xxx
= - {1\over{m\o\sqd r}} \times \bigg [ \cc 
- {1\over{m\sqd\o\uu 4 \cc R\sqd }} \cc \d\prpr(|x| - R)
- {1\over{m\sqd\o\uu 4\cc R\uu 3}}\cc \d\pr(r-R)
\cc \bigg ]
\xxx
= + {1\over{m\uu 3\o\uu 6\cc R\sqd}} \times \bigg [ \cc {1\over r}\cc
\d\prpr(|x| - R) + {1\over{R\cc r}} \cc \d\pr(|x| - R) \cc \bigg ]
\eee

Now differentiate the identity \rr{DeltaFunctionPrimeProductIdentity} 
and subtract $f\pr(r) \cc \d\pr(r - r\ll 0)$ 
to get
\bbb
\hskip-.75in
f(r) \cc \d\prpr(r - r\ll 0) =  f(r\ll 0)\d\prpr(r - r\ll 0) - f\pr(r\ll 0) \cc \d\pr(r - r\ll 0) - f\pr(r) \cc \d\pr(r - r\ll 0)
\xxx
=  f(r\ll 0)\d\prpr(r - r\ll 0) - 2\cc f\pr(r\ll 0) \cc \d\pr(r - r\ll 0) 
+ f\prpr(r\ll 0) \cc \d(r - r\ll 0)\ ,
\eee
so we have
\bbb
 {1\over{R\cc r}} \cc \d\pr(|x| - R) = {1\over{R\sqd}}\cc \d\pr(|x| - R) + {1\over{R\uu 3}}\cc \d(|x| - R)\ ,
 \xxx
 {1\over r}\cc \d\prpr(|x| - R) = {1\over R} \cc \d\prpr(|x| - R) + {2\over{R\sqd}}\cc 
 \d\pr(|x| - R) + {2\over{R\uu 3}} \cc \d(|x| - R)\ ,
\eee
so
\bbb
{1\over r}\cc
\d\prpr(|x| - R) + {1\over{R\cc r}} \cc \d\pr(|x| - R) = 
{1\over R} \cc \d\prpr(|x| - R) + {3\over{R\sqd}}\cc 
 \d\pr(|x| - R) + {3\over{R\uu 3}} \cc \d(|x| - R)\ ,
\eee
so
\bbb
\d\prpr({\bf X}\lrm{cl}) = + {1\over{m\uu 3\o\uu 6\cc R\uu 3}} \times
\bigg [ \cc
 \d\prpr(|x| - R) + {3\over R}\cc 
 \d\pr(|x| - R) + {3\over{R\uu 2}} \cc \d(|x| - R)
\cc\bigg ]
\eee

So, expanded to quadratic order in fluctuations, the operator $\d({\bf X})$ 
is
\bbb
\d({\bf X}) = {1\over{m\o\sqd R}}\cc \d(|x| - R)
 - {1\over{m\sqd\o\uu 4 \cc R\sqd }}  \bigg [ \cc \d\pr(|x| - R) + 
  {1\over R}\cc \d(r-R)   \cc \bigg ]    \cc \widehat{\d{\bf X}}
  \xxx
  + \hh\times  {1\over{m\uu 3\o\uu 6\cc R\uu 3}} \times
\bigg [ \cc
 \d\prpr(|x| - R) + {3\over R}\cc 
 \d\pr(|x| - R) + {3\over{R\uu 2}} \cc \d(|x| - R)
\cc\bigg ] \times \widehat{\d{\bf X}}\sqd
\eee
where the fluctuations $\widehat{\d{\bf X}} = \widehat{\d{\bf X}}(r) $ depend on $r$ as well as $t$ and the angular directions $\O$.  Using the identities
$\rdots$ and $\rdots$ we can write this as an expansion of the
fluctuation fields in normal derivatives at $r=R$:
} 
\begin{eqnarray}
\d({\bf X})
&& = \d(|x| - R) \cc \Bigg [ \cc
{1\over{m\o\sqd R}} - {1\over{m\sqd\o\uu 4 \cc R\uu 3 }} \cc 
 \widehat{\d{\bf X}}
+ {1\over{m\sqd\o\uu 4 \cc R\sqd }}  \cc \widehat{\d{\bf X}}\ll{,r}
+ {3\over{2\cc m\uu 3 \cc \o\uu 6 \cc R\uu 5}}\cc \widehat{\d{\bf X}}\sqd
\nonumber\\
&&\kern+45mm
- {3\over{2\cc m\uu 3 \cc \o\uu 6 \cc R\uu 4}}\cc \pp\ll r \big [ \widehat{\d{\bf X}}\sqd \cc \big ] +  {1\over{2\cc m\uu 3 \cc \o\uu 6 \cc R\uu 4}}\cc \pp\ll r\sqd
\big [ \widehat{\d{\bf X}}\sqd \cc \big ] 
  \cc\Bigg ]\ll{r = R}
  \nonumber\\
  \nonumber\\
  && + \d\pr(|x| - R) \cc \Bigg [ \cc
  - {1\over{m\sqd\o\uu 4 \cc R\sqd }} \cc \widehat{\d {\bf X}}
  + {3\over{2\cc \m\uu 3 \cc \o\uu 6 \cc R\uu 4}}\cc \widehat{\d {\bf X}}\sqd
  - {1\over{m\uu 3\o\uu 6\cc R\uu 3}} \, \pp\ll r \cc \big [ \cc
\widehat{\d{\bf X}}\sqd
\cc\big ] \cc\Bigg ]\ll{r = R}
\nonumber\\
\nonumber\\
&& + \d\prpr(|x| - R) \, {1\over {2\cc m\uu 3\o\uu 6\cc R\uu 3}} \, \big [ \cc{\d{\bf X}}\sqd\cc\big ]\ll{r = R}
 + O\big ( \widehat{\d {\bf X}}\uu 3 \big )\ .
\end{eqnarray}
So, we have traded the abstract expression $\d({\bf X})$ for
an explicit expession in terms of dynamical quantum fields evaluated
at $r=R$, times purely $c$-number distributions in the radial direction,
with calculable coefficients.  

To be sure, we are concerned in this paper with 
simply analyzing the energy of the classical
ground state in the trap, with all bulk and edge operators 
allowed in the action that are consistent with the symmetries.
However, a few comments beyond the present application are useful for context:
\bi
\item{While the above is all at the classical level, this representation of $\d({\bf X})$ is 
well-defined at the quantum level without further information, as long as we are working to first-order in insertions
of the droplet-edge fields.}
\item{Even at classical level, working beyond first order in the coefficients
of the edge-operators requires we resolve the singularity
of the distributions $\d(|x| - R), \d\pr(|x| - R), \cdots$ by giving the $\d$-functions
a finite width  $\sim (m\L)\uu{-\hh}$, where $\L$ is the Wilsonian energy
cutoff satisfying our double hierarchy 
\bbb
E\lrm{IR} \muchlessthan \L \muchlessthan \m = E\lrm{UV}\ .
\eee
}
\item{We can also use this Wilsonian type of analysis to determine the
parametrization of the boundary conditions for $\chi$ at the droplet edge.
Even at semiclassical level, this is useful for a systematic analysis
of the phonon energy spectrum in the ${1/ Q}$ expansion.
Beyond leading order in $Q$, the
parametrization of boundary conditions also has a hierarchical expansion
in negative powers of $Q$; the parameters in the space of boundary 
conditions are related to the parameters in the space of allowed droplet-edge
operators.}
\ei

\subsection{Classical $\m$-scaling and $Q$-scaling of the droplet-edge operators  }\label{ClassicalScaling}

We have now seen that the edge operators have a well-defined operational
meaning beyond the purely classical analysis of the ground state.
In particular, with our understanding of fluctuations under control, we can proceed to 
compute the leading-order effects of various boundary terms in the ${1/ Q}$ expansion
of the energy of the ground state in the harmonic potential, in analogy with the bulk $\m$-scalings 
$\a(\co)$ we defined for bulk operators in the harmonic potential.

\subsubsection{$\m$-scalings of (unintegrated) dressed edge operators}

Using our Jacobian \rr{OperatorDeltaFunction},\rr{JacobianFormulaClassicalApprox}, we have
\bbb
\d({\bf X}) = \big | \cc {{d{\bf X}}\over{dr}} \cc \big |\uu{-1} \cc \d(|x| - R)
= {\bf Y}\uu{-\hh}\cc \d(|x| - R)\ ,
\een{MeasureFactorDeltaFunctionOfBoldX}
plus fluctuation terms, which we ignore when computing the classical
ground state energy.  

To be clear, the scaling behavior of edge operators is distinct from bulk operators,
insofar as we are defining the $\m$-scaling of the former
by their behavior as we scale up $\m$ and evaluate the operator $\co(x)$
at a point on the edge $|x| = R = {{\sqrt{2\m}}/({\o\sqrt{m}})}$, 
which itself grows with $\m$.  That is, we can assign operators a
boundary $\m$-scaling exponent $\b(\co)$ that can be different
from its bulk $\m$-scaling exponent.
For instance, the operator ${\bf Y}$ in the trap background
is proportional to $m\sqd \o\uu 4 r\sqd$, so at fixed $r$ it scales
as $\m\uu 0$.  However, the bulk $\m$-scaling exponent involves
evaluating ${\bf Y}$ at $|x| = R = {{\sqrt{2\m}}/({\o\sqrt{m}})}$, which
means the boundary $\m$-scaling exponent of ${\bf Y}$ is $+1$,
\bbb
\b({\bf Y}) = +1\ .
\eee
As for ${\bf Z}$, its classical value is simply proportional to
$m\o\sqd$, independent of position, so its boundary $\m$-scaling
exponent vanishes (as does its bulk $\m$-scaling exponent):
\bbb
\b({\bf Z}) = 0\ .
\eee

\subsubsection{$\m$-scalings of integrated edge terms}\label{MuScalingsOfIntegratedEdgeTermsGeneralDCalculation}

When integrating an edge operator, we must first 
take care to estimate the $\m$-scaling of the $\d$-function contribution $\d({\bf X})$.  
As noted above in eqn.~\rr{MeasureFactorDeltaFunctionOfBoldX}, the
translation of $\d({\bf X})$ into a $\d$-function of radial position
comes with an extra measure factor of ${\bf Y}\uu{-\hh}$, 
\bbb
\d({\bf X})  = {\bf Y}\uu{-\hh}\cc \d(|x| - R)\ ,
\een{MeasureFactorDeltaFunctionOfBoldXRECAP}
so
\bbb
\b\big [ \cc \d({\bf X}) \cc \big ] = -\hh\cc \b({\bf Y}) = -\hh\ .
\eee

Generally, edge operators are integrated over a $(d-1)$-dimensional sphere
of radius $R$, and $R$ itself scales as $\m\uu{\hh}$ \rr{DefOfDropletRadius}.  If we introduce the operation ${\cal I}$, 
which promotes an undressed operator to its dressed counterpart and performs
the appropriate integration, we can compactly encode the sequence of manipulations needed to 
extract the $\mu$-scalings of interest for edge 
operators. 
Namely, following the edge-operator decomposition in eqn.~\rr{EdgeOperatorGeneralForm}, we obtain
\bbb
{\cal I}[\co\lrm{{undressed}} ]\equiv
\int\cc d\uu d x \cc \co\lrm{edge} = \int \cc d\uu d x \cc \d({\bf X}) \cc (m\sqd {\bf Y})\uu{{{d+4 - \D\lrm{und}}\over 6}}\co\lrm{undressed}\ ,
\eee
which has the classical value
\begin{eqnarray}
\big\langle {\cal I}[\co\lrm{{undressed}} ]  \big \rangle &=& {\cal A}\ll{d-1} \cc R\uu{d-1}   (m\sqd {\bf Y})\uu{{{d+1 - \D\lrm{und} }\over 6}}\, \langle\co\lrm{undressed}\rangle 
\nonumber\\
\nonumber\\
& \propto & \m\uu{{{2d-1}\over 3} - {{\D\lrm{und}}\over 6}} \, \langle\co\lrm{undressed}\rangle \ .
\end{eqnarray}
The brackets in the equation above denote the classical value of the undressed operator
evaluated at the boundary of the droplet edge in the classical ground
state solution in the harmonic trap.  If the $\m$-scaling of 
the undressed operator is
\bbb
\langle\co\lrm{undressed}\rangle \sim \m\uu{\b\lrm{und}}\ ,
\eee
then the integrated and dressed operator scales as 
\bbb
\langle {\cal I}[\co\lrm{{undressed}} ] \rangle \sim \m\uu{\g\lrm{total}} \equiv \m\uu{{{2d-1}\over 3} + \g\lrm{term}}\ ,
\label{GammaTotalGeneralRepresentation}
\eee
where we define the exponents to be
\bbb
\g\lrm{term}\equiv \b\lrm{und} - {1\over 6}\D\lrm{und}\ ,
\xxnn
\g\lrm{total} \equiv {{2d-1}\over 3} +\sum\ll a \g\ll{{\rm term~}a}\ .
\een{GammaTotalFormula}
The exponent $\g\lrm{term}$ can be thought of as the effective $\m$-scaling exponent
of the dressed operator $(m\sqd {\bf Y})\uu{- {1\over 6}\D\lrm{und}}
\co\lrm{undressed}$, which is just the undressed operator with
a power of the boundary dressing appended to bring the dressed operator
to zero scaling dimension.

This recipe for constructing conformal edge operators and computing
their scaling at large quantum number is essentially identical to that of the analysis for
bulk \cite{Hellerman:2014cba} and boundary \cite{Hellerman:2016hnf} operators in effective string theory.  
As discussed above, and now made explicitly clear, the one difference is that the 
position of the edge is operator-valued rather than fixed, and so instead
of a delta function of a fixed coordinate position, as in \cite{Hellerman:2014cba}, we have an operator-valued $\d$-function $\d({\bf X})$ to
restrict the support of the operator in the effective theory of the finite
droplet in NRCFT. 

\subsubsection{Classification of Lagrangian perturbations by their classical $\m$-scaling}
\label{MuScalingsOfIntegratedEdgeTermsGeneralDFirstTryCutAndPastedFromNotes}

To classify droplet-edge terms in the action by their $Q$-scaling
(or equivalently their $\m$-scaling), we would
like to show that there are only a finite number of undressed operators $\co\lrm{und}$ with $\b\lrm{und} - {{\D\lrm{und}}\over 6}$ greater than any given number, where $\D\lrm{und}$ and $\b\lrm{und}$ are the conformal dimension
of $\co\lrm{und}$ and its boundary $\m$-scaling, respectively.
Let us start by considering what type of undressed $\co\lrm{und}$ we
can make from the derivatives of ${\bf X}$.  To every undressed
operator $\co\lrm{und}$, we assign the combination $\g\equiv \b\lrm{und} - {1\over 6} \D\lrm{und}$.

\begin{table}[h!]
\begin{center}
\caption{Scaling decomposition of some simple objects.}
\label{ScalingTableVevfulObjects}
\begin{tabular}{ c||c|c|c } 
 \toprule
$~~{\cal O}\lrm{undressed}~~$
     & $~~\b\lrm{und}~~ $ & $~~\D\lrm{und}~~$ & $\g\lrm{term}\equiv \b\lrm{und} - {1\over 6}\D\lrm{und}$   \\ \hline\hline
 $1$ &  $0$ & $0$  & $0$     \\ \hline
 $\pp\ll x{\bf X}$ & $+\hh$ & $+3$ &$0$   \\ \hline
 $(\pp\ll x)\sqd {\bf X}$ & $0$ & $+4$ & $-{2\over 3}$   \\ 
\end{tabular}
\end{center}
\end{table}
It helps to start with a table of some simple objects (see Tab.~\ref{ScalingTableVevfulObjects}). 
In this table we have included only objects with nonzero vevs in the classical vacuum solution, so that
their classical $\m$-scaling exponent $\b\lrm{und}$ is well-defined.  Note that nothing contributes with 
positive $\g_{\rm term}$.

It is especially worth noting what {\it does not} appear in Table~\ref{ScalingTableVevfulObjects}.  
We have omitted, for instance,  $\pp\ll t {\bf X}$ and $\pp\ll x\uu\ell {\bf X}$ for $\ell \geq 3$, 
because these vanish classically.
We have also omitted background terms proportional to the spatial magnetic field ${\bf F}\ll{ij}$ and its derivatives,
because the magnetic field is zero in the background
defining the harmonic trap.  Background terms proportional to the electric field and
its derivatives can also be left off.  The electric field ${\bf E}\ll i$ is nonvanishing in the harmonic trap, but the spatial gradient ${\bf X}\ll{,i}$ of the
classical solution ${\bf X}$ is equal to ${\bf E}\ll i$, so we have the relation ${\bf E}\ll i =  {\bf X}\ll{,i} + ({\rm vevless~operators})$.
Finally, we also omit spatial gradients of $\chi$.  The term $\chi\ll{,i}$ itself is not gauge-invariant, and
its gauge-invariant completion $\gg\ll i \chi = {\bf F}\ll{ij} $ vanishes identically in this background.  We summarize these omissions in Table \ref{TableOmittedObjectsAndReasons}.
\begin{table}[h!]
\begin{center}
\caption{Omitted objects.}
\label{TableOmittedObjectsAndReasons}
\begin{tabular}{ c|c } 
 \toprule
   $~~{\cal O}~~$ &  Classical vacuum solution of the harmonic trap:   \\ \hline\hline
 $\pp_t{{\bf X}}$    &    ${\bf X}$ is constant in time     \\ \hline 
 $\pp\ll x \uu \ell{{\bf X}}, l\geq 3 $   &    ${\bf X}$ is  quadratic in $x$  \\ \hline
  ${\bf F}\ll {ij}$    &   vanishes in this background  \\ \hline
  ${\bf E}\ll i$ &    ${\bf E}\ll i =  {\bf X}\ll{,i} + ({\rm vevless~operators})$      \\ \hline
 $\gg\ll {i\ll i}\cdots \gg\ll{i\ll k} \chi$   &  zero because $\gg\ll i \chi = {\bf F}\ll{ij} = 0$   \\ 
\end{tabular}
\end{center}
\end{table}

So, put simply, the total $\mu$-scaling exponent of dressed edge operators is
\bbb
\g\lrm{total} \equiv {{2d-1}\over 3} + \sum\ll a \g\ll a = {{2d-1}\over 3} - {2\over 3}\#({\bf X}\ll{,ij}) = {1\over 3}\bigg ( \cc 2[d - \#({\bf X}\ll{,ij})] -1\bigg ) \ .
\een{SimpleFormulaForMuScalingOfIntegratedBoundaryOperator}
Of course, we can translate this into a $Q$-scaling, since
$\m \propto Q\uu{1\over d}$ (by eqn.~\rr{SimplyExpressedRelationshipBetweenMuAndQInTheHarmonicTrapRECAPFirstAppearance}), 
so the $Q$-scaling of integrated edge terms is
\bbb
{\cal I}\bigg [ \cc \prod\ll a \cc \co\upp a\lrm{und} \cc \bigg ] 
 \propto Q\uu{ {{\g\lrm{total}\over d}}}\ .
\een{ExplicitRepresentationOfIntegratedGenericEdgeOperator}
Per eqn.~\rr{SimpleFormulaForMuScalingOfIntegratedBoundaryOperator}, the important arithmetic is contained in the number of ${\bf X}\ll{,ij}$
in the operator, and, in particular, the only droplet-edge operators with nonnegative $\m$-scaling at the classical level are those with 
\bbb
\#({\bf X}\ll{,ij}) < d\ .
\een{CriterionForNonnegativeMuScalingAtClassicalLevel}

Note that only a finite number of operators exist with scaling exponent
$\g\lrm{total}$ greater than a specified floor $\g\lrm{min}$.
In particular, the classification above reduces the set of independent Wilson coefficients contributing classically at a given order
to a finite basis.  Any operator with a nonzero
classical value can be built out of the three operators
in Table~\ref{ScalingTableVevfulObjects}, modulo operators that
vanish classically.  The identity is trivial algebraically, so
we only have to count powers of ${\bf X}\ll{,i}$ and ${\bf X}\ll{,ij}$.
However, any pair of ${\bf X}\ll{,i}$ in which the vector indices
are contracted with each other gives a factor of ${\bf Y}$,
which is effectively trivial because it is absorbed by
the ${\bf Y}$-dressing.  The only way to make algebraically
independent boundary operators with nonzero
classical values in the ground state  
configuration is then to include powers of ${\bf X}\ll{,ij}$, contracting them either with the metric, with ${\bf X}\ll{,i}$, or with each other.  Each power
of ${\bf X}\ll{,ij}$ contained in the numerator, has a $\g$-contribution
of $\g = -{2\over 3}$.  So there are only a finite number of terms
that can be constructed with the total $\g$ exponent greater than 
any fixed amount: Any lower limit on $\g\lrm{total}$ sets an
upper limit on the number of ${\bf X}\ll{,ij}$ that can be included
in a term.  With any fixed number of ${\bf X}\ll{,ij}$, there are
only a finite number of invariant terms that can be
constructed.

\subsection{A sufficient condition for allowed edge operators}

So far in this section we have performed a coarse analysis of scaling laws for edge operators, with the aim of bounding the number of operators that can appear at a given order.
We have not so far given any sort of \emm{lower} bound for the number of independent operators with a given \bbd{\m}-scaling at any order, and we have not explicitly constructed any
specific operators that are allowed, beyond the simplest example $m\uu{-2} \d({\bf X}) (m\sqd {\bf Y})\uu{ {{d+4}\over 6}}$.  It is of course important to be sure that one can reliably construct \rwa{all} edge operators at a given order; the derivation of sum rules and other relations among terms in the large-\bbd{Q}
expansion of observables, depends on knowing the number of independent Wilson coefficients at a given order, including edge Wilson coefficients. 

 We do not attempt it here because of
certain subtleties involved in the criteria for gauge invariance and conformal invariance at the edge.  The naive criteria for invariance are plausibly too strict in the presence of the \bbd{\d}-function
that appears in edge terms.  For instance, there may be local operators on which a conformal lowering operator produces a term proportional to an undifferentiated ${\bf X}$.  Such a term would
not be primary in the bulk, but is primary when dressed with a \bbd{\d({\bf X})} and turned into an edge operator.  Due to these and related subtleties, we defer a fuller analysis of sufficient conditions
for allowed edge operators to future work.

However there is one recipe for constructing invariant edge operators that already generates examples contributing certain \bbd{\m-}scalings to the energy, that appear to have gone unnoticed in the literature so far.  Since bulk operators are already gauge-invariant and conformally invariant, we can always start with a bulk operator
and construct an edge operator by the following recipe:
\bii
\item{Start with a bulk term in bipartite form \rr{XDressingRuleForBulkOperators}, $\co\lrm{bulk} = \co\lrm{undressed} \cc {\bf X}\uu{1 + {d\over 2} - \hh \D\lrm{und}}$.  With no loss of generality we can assume \bbd{\co\lrm{undressed}} contains no undifferentiated powers of \bbd{{\bf X}}.}
\item{Strip off the ${\bf X}$-dressing, to isolate the numerator  \bbd{\co\lrm{undressed}}.  The key point is that  \bbd{\co\lrm{undressed}} must be a gauge-invariant conformal primary, since
\bbd{{\bf X}} is a conformal primary, and terms in the Lagrangian density must be gauge-invariant and primary as well. }
\item{Then, re-dress the numerator  \bbd{\co\lrm{undressed}} as an edge term by appending the edge dressing $ \d({\bf X})( {\bf Y}/m)\uu{{1\over 6}(d + 4 -\D\lrm{und})}$, to obtain a new edge
operator 
\bbb
\co\lrm{edge} =  \d({\bf X})( {\bf Y}/m)\uu{{1\over 6}(d + 4 -\D\lrm{und})} \co\lrm{undressed}\ .
\een{NewEdgeOp}
}
\item{The dressing factor $( {\bf Y}/m)\uu{{1\over 6}(d + 4 -\D\lrm{und})} $ is a conformal primary of weight $d+4 - \D\lrm{und}$
and the $\d$-function factor $\d({\bf X})$ is a conformal primary
of weight $-2$, so the new edge operator \rr{NewEdgeOp} is
a conformal primary of weight $2+d$, and is an
allowed term in the action.}
\ei
Again we emphasize this construction may not necessarily
produce \rwa{all} possible edge terms, but it is a concrete
construction that produces some examples of edge terms
that might not be noticed otherwise, and indeed do not
seem to have shown up in the literature so far.

In particular we can apply this construction to the bulk
term ${\cal L}\ll{c\ll 2} \propto m\uu{\hh(d-2)}\cc {\bf X}\uu{{{d\over 2}} - 1}\cc{\bf Z}$ showing up as a term \rr{NLOTerm2} in the bulk action.  Here we have $\co\lrm{undressed} = m\uu{\hh(d-2)}\cc{\bf Z}$ so and $\D\lrm{und}= 4$, so the corresponding
edge operator is
\bbb
\co\lrm{edge} = m\uu{{{d-2}\over 2}}\cc \d({\bf X}) \cc ({\bf Y} / m)\uu{{d\over 6}} \cc {\bf Z}
= m\uu{-1}\cc \d({\bf X}) \cc (m\sqd {\bf Y} )\uu{{d\over 6}} \cc {\bf Z}
\eee
The undressed term ${\bf Z}$ has $\m$-scaling exponent $\b\lrm{und} = 0$ and a nonzero expectation value in the classical ground-state solution; so by formula \rr{GammaTotalFormula} the integral of the dressed edge $\co\lrm{edge}$ has $\m$-scaling $\g\lrm{total} = {{2d-3}\over 3}$.  We compute its contribution in sec. \ref{ContributionOfDressedZEdgeOperator}.

In this section we have analyzed the $\m$-scalings of edge operators at the classical level in the EFT, reducing to a finite problem the
enumeration of edge operators that contribute classically with $\m$-scaling above any given exponent in any given dimension $d$.
This analysis is
incomplete without an analysis of the quantum contributions of operators which vanish classically in the large-charge vacuum, both tree-level contributions at second and higher order
in perturbation theory, and loop contributions.  We will give a loose upper bound on the quantum scalings of vevless operators in the next section, that will 
suffice to make the counting of terms that may contribute quantum mechanically, into a finite problem.

\section{Energies in the harmonic trap at NLO}\label{NLOTrapEnergies}

\subsection{Energy shift from the Lagrangian perturbation}

To simplify the overall treatment, we focus here (mostly) on 
the $c\ll 1$ term in the NLO bulk Lagrangian, which is sufficient to illustrate the issues of
principle involved in the renormalization of near-edge singularities
with boundary counterterms.  The integrated $c\ll 2$ term in the Lagrangian is convergent in $d=2$, and for completeness we discuss
it briefly in section \ref{EffectOfTheOtherBulkTermInTwoDimensionsAtLeast}.

The $c\ll 1$ term in the NLO bulk lagrangian, given in eqs.\rr{NLOStructureHere}, \rr{YDef}, and \rr{NLOTerm1}, is
\bbb
{\cal L}\ni m\uu{{{d-2}\over 2}}\, c\ll 1 \,
{\bf X}\uu{{d\over 2} - 2}\, (\pp\ll i {\bf X}\cc \pp\uu i {\bf X})\ .
\een{secondorderLeffC1TermRecap}
To evaluate its effect on the energy of the charged ground state
in the trapping potential, we now recall a simple lemma from
classical mechanics:  For a charged ground state of a classical system
with a global symmetry, the first-order perturbation of the energy is
just the negative of the perturbing Lagrangian, evaluated in the
(unperturbed) ground state.
This lemma generalizes the usual relationship between
the first-order perturbing Lagrangian and first-order perturbing
Hamiltonian for the overall ground state.  In fact, we can reduce
the more general case to the case of the overall ground state
by adding an explicit chemical potential to the action.

For the case at hand, if $c\ll 1$ is a coupling constant multiplying a small
term in the Lagrangian, then
\bbb
E(Q)\cc\big |\ll{O(c\ll 1)} = - L(\m)\cc \big |\ll{O(c\ll 1), ~\m\to \m\ll 0(Q)}\ ,
\een{FirstOrderPerturbationFormula}
where $\m\ll 0(Q)$ is the expression for the chemical potential
evaluated at charge $Q$ in the unperturbed action at $c\ll 1 = 0$.
According to the lemma above, we have
\begin{eqnarray}
(\Delta E)\ls {c\ll 1} &=& H\ls{c\ll 1} = - L\ls{c\ll 1} = - \int \cc d\uu d x \cc {\cal L}\ls{c\ll 1}
\nonumber\\
&=& -m\uu{{{d-2}\over 2}}\, c\ll 1 \, \int \cc d\uu d x \cc 
{\bf X}\uu{{d\over 2} - 2}\, (\pp\ll i {\bf X}\cc \pp\uu i {\bf X})\ ,
\label{EnergyShiftEqualsNegativeLagrangianPerturbation}
\end{eqnarray}
where ${\bf X}$ is evaluated in the unperturbed 
classical solution (i.e., the $c\ll 1 = 0$ classical solution).
The energy shift thus appears as
\bbb
(\Delta E)\ls {c\ll 1} = - 
m\uu{{{d+2}\over 2}}\, c\ll 1 \,  
\o\uu 4\int \cc d\uu d x \, r\sqd \,
\left( \m - {{m\o\sqd}\over 2} \cc r\sqd \right)\uu{{d\over 2} - 2}\ .
\een{SHCalculationOfEnergyShiftAsAnIntegralGeneralDimensionD}

Evaluating the integral in angular variables, using formula \rr{UnitSphereAreaValueRECAP0}
for the area of the unit $d-1$ sphere, and formula \rr{DefOfDropletRadius} for
the classical size $R$ of the droplet, we can straightforwardly evaluate the integral in
$d$ spatial dimensions:
\bbb
(\Delta E)\ls {c\ll 1} = 
- 4\pi\m\, c\ll 1\, \left( \sqrt{2\pi}\cc {{\m\over\o}}\right)\uu{d-2} 
\left(  {{\G({d\over 2} + 1)  \cc \G({d\over 2} - 1)}\over{\G({d\over 2})\cc\G(d) }}\right)\ .
\een{SHCalculationOfEnergyShiftAsAnIntegralGeneralDimensionDIntermediateStage4}
We would also like to express this in terms of the total charge $Q$ in the trap.  Using the relationship \rr{SimplyExpressedRelationshipBetweenMuAndQInTheHarmonicTrapRECAPFirstAppearance},
we have
\bbb
\m\uu{d-1} =  \xi\uu{{{d-1}\over 2}} \,
\left( {{\G(d+1)}\over 2}  \right)\uu{{{d-1}\over d}} \o\uu{d-1}\cc Q\uu{{{d-1}\over d}}+ O(c\ll 1\cc Q\uu{{{d-3}\over d}})\ ,
\eee
so, in terms of $Q$,
\bbb
(\Delta E)\ls {c\ll 1\uu 1,{\rm ~fixed~}Q} = 
-  2c\ll 1\o \cc(2\pi)\uu{{d\over 2}}\cc \xi\uu{{{d-1}\over 2}} 
\left( {{\G(d+1)}\over 2}  \right)\uu{{{d-1}\over d}}   {{\G({d\over 2} + 1)  \cc \G({d\over 2} - 1)}\over{\G({d\over 2})\cc\G(d) }}  
  \, Q\uu{{{d-1}\over d}} \ . 
\een{SHCalculationOfEnergyShiftAsAnIntegralGeneralDimensionDIntermediateStage4ExpressedInTermsOfQ}
Alternatively, in terms of $c\ll 0$,
\bbb
(\Delta E)\ls {c\ll 1\uu 1,{\rm ~fixed~}Q} = 
-  2^{3 - {2\over d}}\cc \frac{\sqrt{2\pi}\cc c\ll 1\, \o}{c\ll 0\uu{{{(d-1)}/d}}}  \cc d\uu{{1\over d}}\cc (d+2)\uu{-{{d-1}\over d}}
\cc  {{\G({d\over 2} + 1)  \cc \G({d\over 2} - 1)}\over{[\G({d\over 2})]\uu{2 - {1\over d}}\cc[\G(d+1)]\uu{{1\over d}} }}  
 \, Q\uu{{{d-1}\over d}} \ . 
\een{SHCalculationOfEnergyShiftAsAnIntegralGeneralDimensionDIntermediateStage4ExpressedInTermsOfQAndC0}

This calculation is convergent in dimension $d\geq 3$.  For $d = 3$ in particular, we have
\bbb
(\Delta E)\ls {c\ll 1}\cc \biggl |\ll{d = 3}  = 
- 4\pi\, c\ll 1\, \sqrt{2\pi}\cc {{\m^2\over\o}}\,  {{\G({5\over 2})  \cc \G(\hh)}\over{\G({3\over 2})\cc\G(3) }}
= -3 \sqrt{2} \cc \pi\sqd \cc {{c\ll 1\m\sqd}\over\o}\ . 
\een{SHCalculationOfEnergyShiftAsAnIntegralGeneralDimensionDIntermediateStage4EvaluatedInThreeSpatialDimensions}
Or, in terms of $Q$,
\bbb
\m\cc \big |\ll{d=3,~{\rm leading~order}}  = \xi\uu{\hh} \,
\left( 3 \cc Q \right)\uu{{1\over 3}}\cc\o\ ,
\eee
so
\bbb
(\Delta E)\ls {c\ll 1}\cc \biggl |\ll{d = 3}  = - 3\uu{{5\over 3}}\cc \sqrt{2} \cc \pi\sqd\cc c\ll 1\cc \xi\o\cc Q\uu{{2\over 3}}\ .
\eee
This agrees with the $c\ll 1\uu 1$ term in eqn.~(9) of \cite{Son:2005rv} (with $\o_1 = \o_2 = \o_3 = \o$).

In $d=2$ the expression \rr{SHCalculationOfEnergyShiftAsAnIntegralGeneralDimensionDIntermediateStage4ExpressedInTermsOfQAndC0} is 
divergent, so we have to regulate and renormalize it.  
In the following several sub-sections we will deal with this computation.
In Sec.~\ref{ConformalSharpCutoff} we do so with a sharp cutoff in $d=2$.  
Then, in \ref{DimRegPartA},  \ref{EdgeCountertermIdentification}, and \ref{DimRegPartC}, we will 
employ dimensional regualrization to cross-check the result.  Namely, we derive the universal
coefficient of the $\m\cc {\log}(\m/\o)$ term in the energy,
which in terms of the particle-number charge $Q$ corresponds to a term of
order $Q\uu{\hh}\, {\log}(Q)$.

\subsection{Conformal sharp cutoff in $d=2$}\label{ConformalSharpCutoff}

Let us first calculate the energy in $d=2$ by cutting off
the integral explicitly near the droplet edge.  It will turn out that the sharp conformal cutoff is 
equivalent to the ``cloud radius" cutoff used in
\cite{Son:2005rv,Kravec:2018qnu}, but organized in a more manifestly conformal way.

The simplest conformally invariant way to regulate the integral is to specify the cutoff ${\bf X} = \e \cc( {\bf Y} /m)\uu{{1\over 3}}$.  
With this definition of the cutoff, the parameter $\e$ is dimensionless
in $\hbar = 1, m\neq 1$ dimensional analysis.  
Since both ${\bf X}$ and ${\bf Y}$ are primary fields, the cutoff parameter $\e$ is actually conformally invariant.

Note that the integral diverges if taken in the limit of fixed $\m$ and $\e\to 0$.
This is not how the limit should ever be taken, though; one should
always regulate and renormalize at fixed $\e$ and take $\m$
large.  At any fixed $\e$, the loop corrections and higher-derivative
corrections are suppressed by powers of $\m$, even near
the edge.   This is similar to the situation in effective string theory, 
in which the derivative expansion of the worldsheet EFT is reorganized 
at the boundary of an open string with freely-moving endpoints \cite{Hellerman:2016hnf}.
In these cases, the EFT itself does not actually break down, but some other invariant 
takes over as the dressing operator appearing in denominators of effective terms in the 
singular region: The dressing rule at the boundary (or defect, or droplet edge) is not the 
same as the dressing rule in the bulk, but there is still a well-defined dressing rule and a 
well-defined derivative expansion that generates a perturbative expansion at large quantum number.

In the case of droplet-edge operators, the dressing field at the edge is ${\bf Y}$.   
So it is only the expansion in $\pp / {\bf X}\uu\hh$ that
breaks down near the droplet edge, not the derivative expansion
of the EFT altogether.  The low-energy expansion is reorganized
into a derivative expansion in $\pp / {\bf Y}\uu{{1\over 6}}$. 
This fact is important when we come to the point of fully classifying operators at the droplet edge.

In the classical solution, the formula for the cutoff point, is
\bbb
R\ll \e = R - \d\ll \e\ ,
\xxx
\d\ll \e \simeq 2\uu{- {1\over 6}}\, m\uu{- {1\over 2}}\,
\o\uu{- {1\over 3}}\,\m\uu{-{1\over 6}}\,  \e \ ,
\eee
where the error is of $O(\e\sqd\cc m\uu{-\hh}\cc \o\uu{{1\over 3}} \cc \m\uu{-{5\over 6}})$.

Cutting off the integral at the point $r = R - \d\ll\e$ (and working in $d=2$), we get
\begin{eqnarray}
L\ls{c\ll 1}\to \int\ll{r < R - \d\ll\e} \cc d\uu d x\cc {\cal L}\lrm{c\ll 1,~\m} &=& \area{d-1}\cc \int
\ll 0\uu{R - \d\ll\e} \cc dr\cc r\uu{d-1}\cc  {\cal L}\lrm{c\ll 1,~\m}
\nonumber\\ \nonumber\\
&=& 2\pi \cc \int
\ll 0\uu{R - \d\ll\e} \cc dr\cc r\cc  {\cal L}\lrm{c\ll 1,~\m}
\nonumber\\ \nonumber\\
&=& - 4\pi c\ll 1 \m\cc \log (\d / R) + O(\m\uu 1)
\nonumber\\ \nonumber\\
&=&  {{8\pi c\ll 1 \m}\over 3} \cc \log \left( {\m\over{\o\cc \e\uu{3\over 2}}} \right) + O(\m\uu 1)\ .
\label{BareLagrangianShiftSharpCutoffTwoDimensions}
\end{eqnarray}
The nonlogarithmic term is scheme-dependent and not calculable within
the EFT.  Rather, its coefficient can be absorbed into a local edge counterterm, as we shall see in sec. \ref{EdgeCountertermIdentification}.

The leading-order relationship between chemical potential $\m$ and
charge $Q$ in the isotropic harmonic trap with frequency $\o$, is
\bbb
\big ( \cc {\m\over\o} \cc \big ) = \xi\uu{\hh} \, Q\uu{\hh} \ ,
\een{SimplyExpressedRelationshipBetweenMuAndQInTheHarmonicTrapRECAPFirstAppearanceRestrictedToTwoDimensions}
or equivalently
\bbb
Q=  \xi\uu{-1}
\, \big ( \cc {\m\over\o} \cc \big )\sqd \ ,
\een{SimplyExpressedInverseRelationshipBetweenMuAndQInTheHarmonicTrapRECAPFirstAppearanceRestrictedToTwoDimensions}
which are taken from \rr{SimplyExpressedInverseRelationshipBetweenMuAndQInTheHarmonicTrapRECAPFirstAppearance},\rr{SimplyExpressedRelationshipBetweenMuAndQInTheHarmonicTrapRECAPFirstAppearance}, and evaluated in $d=2$.
Substituting in these leading-order relations,
we have
\begin{eqnarray}
(\Delta E)\lrm{c\ll 1,\cc Q} &=& - (\Delta L)\lrm{c\ll 1, \cc \m} \cc \bigg |\lrm{\m\to Q\ll 0(\m)}
\nonumber\\ \nonumber\\
&=& - {{8\pi c\ll 1}\over 3} \xi\uu{\hh}  Q\uu{\hh}\, \o \cc \log \left( {{\xi\uu{\hh}  Q\uu{\hh}}\over{\e\uu{3\over 2}}} \right) + O(Q\uu\hh \o)
\label{FinalAnswerOrderC1EnergyShiftInSharpCutoff}
\end{eqnarray}
Note that we employ here the mechanics lemma introduced above.

Now we will 
cancel the divergence with a counterterm to derive the renormalized energy.  This discussion is essentially equivalent 
to the derivation in \cite{Kravec:2018qnu}, the only difference being an emphasis on doing the regularization and renormalization
with manifest conformal covariance.

In $d=2$, the operator $\d({\bf X}){\bf Y}$ is dimension $4$ and its integral scales 
as $\m\uu{1}$, according to formulas \rr{GammaTotalGeneralRepresentation}, \rr{GammaTotalFormula}, in the case where the 
undressed operator is the identity. Adding this term with a coefficient 
proportional to $c\ll 1  {\log}(\e\uu{-{3\over 2}})$ cancels the $\e$-dependence of the 
bare term \rr{FinalAnswerOrderC1EnergyShiftInSharpCutoff}, leaving a cutoff-independent 
result: \rr{BareLagrangianShiftSharpCutoffTwoDimensions}:
\bbb
(\Delta E)\lrm{c\ll 1,\cc Q,~{\rm renormalized}}
= - {{4\pi c\ll 1}\over 3} \, \xi\uu{\hh}  \, Q\uu{\hh} \, \o \cc {\log} \left( \xi Q \cc \right) + O(Q\uu\hh \o) \ .
\een{FinalAnswerOrderC1EnergyShiftInSharpCutoffRenormalized}
We will now go on to recalculate this answer in dimensional regularization, 
checking that we get the same result for the coefficient of $Q\uu{\hh} {\log}(Q)$ in the operator dimension in $d=2$.
Note that there can be no conformal boundary counterterm with a logarithmic dependence on $\m$ (since the argument of the logarithm must be dimensionless, and there is no dimensionful
parameter avilable), so we expect the 
coefficient of the $Q\cc {\log}(Q)$ term to be universal and scheme-independent, given the value of the bulk 
coefficient $c\ll 1$.  We shall now check this expectation by calculating the same contribution to the energy in dimensional regularization.

\subsection{Evaluation of the bare energy at order $c\ll 1\uu 1$ in dimensional regularization, near $d=2$}\label{DimRegPartA}

Let us return to eq. \rr{SHCalculationOfEnergyShiftAsAnIntegralGeneralDimensionDIntermediateStage4}, which we recap here
for convenient reference,
\bbb
(\Delta E)\ls {c\ll 1} = 
- 4\pi\m\, c\ll 1\, \left( \sqrt{2\pi}\cc {{\m\over\o}} \right)\uu{d-2} 
\, \left(  {{\G({d\over 2} + 1)  \cc \G({d\over 2} - 1)}\over{\G({d\over 2})\cc\G(d) }}
\right)\ ,
\een{SHCalculationOfEnergyShiftAsAnIntegralGeneralDimensionDIntermediateStage4Recap}
and expand the expression near $d=2$.
Namely, we find
\bbb
(\Delta E)\ls {c\ll 1} = -{{8\pi c\ll 1\m}\over{d-2}} - 8\pi c\ll 1\cc \m\cc  {\log}\left(  {{\m\over\o}}\right)+ O\left[\m\uu 1 \cc (d-2)\uu 0\right] 
+ O\left[(d-2)\uu 1\right]\ . 
\een{SHCalculationOfEnergyShiftAsAnIntegralGeneralDimensionDIntermediateStage5}
There is a divergence proportional ${1/ {(d-2)}}$ with
coefficient of order $\m\uu{1}$.  We will now
see that this divergence is an ultraviolet
divergence corresponding to a local boundary
counterterm. 

\subsection{Identification and coefficient of the boundary counterterm in dimensional regularization, at $d=2$}\label{EdgeCountertermIdentification}

To understand the form of the counterterm, we refer to Sec.~\ref{DropletEdgeOps} and consult the results
of the boundary operator analysis therein.  
In eqns.~\rr{SimpleFormulaForMuScalingOfIntegratedBoundaryOperator} and \rr{CriterionForNonnegativeMuScalingAtClassicalLevel}, we
classified all possible boundary
operators that could contribute classically at order $\m\uu 0$ or larger, which of course should include 
any possible counterterm to cancel
the divergence in expression \rr{SHCalculationOfEnergyShiftAsAnIntegralGeneralDimensionDIntermediateStage5}.  
By formula \rr{SimpleFormulaForMuScalingOfIntegratedBoundaryOperator}, an edge operator scaling as $\m\uu{1}$ in $d=2$ must have
no ${\bf X}\ll{,ij}$ appearing within.  
The only available scalar operator is then the dressed identity, since additional 
powers of $(\pp {\bf X})\sqd = {\bf Y}$ are, by definition, cancelled by 
the ${\bf Y}$-dressing to adjust the conformal dimension
to marginality.  So the only available counterterm is a multiple of $\d({\bf X})\cc {\bf Y}\uu{+1}$:
\bbb
\Delta H\lrm{edge} \ni \k\ll d \cc {\cal I}\big[1 \big] \ ,
\eee
with
\bbb
{\cal I}\big [ 1 \big ] 
\equiv m\uu{-2}
\int \cc d\uu d x \cc \d({\bf X}) \, (m\sqd\cc {\bf Y})\uu{{d+4}\over 6} 
\een{ExplicitRepresentationOfIntegratedEdgeOperatorIdentity}
Now we can evaluate this integral, again using dimensional regularization.

Recalling from above that
\bbb
\d({\bf X}) = |\pp\ll r \cc{\bf X}|\uu{-1}\cc \d(|x| - R) = {\bf Y}\uu{-\hh}\, \d(|x| - R)\ ,
\eee
the integrated edge operator ${\cal I}[1]$ is thus
\bbb
{\cal I}\big [ \cc 1 \cc \big ]  =m\uu{{{d-2}\over 3}}\cc R\uu{d-1} \, \area{d-1}\, {\bf Y}\uu{{{d+1}\over 6} } \cc\bigg |\ll{r=R} \ ,
\een{IntegratedDressedIdentityOperatorEvaluatedInClassicalSolutionAtFixedMu}
where ${\bf Y}$ is evaluated at $r=R = {{\sqrt{2\m}}/({\o\sqrt{m}})}$.
Then, using the classical value \rr{BoldYInvariantClassicalAtDropletEdgeValue} of the ${\bf Y}$ invariant at $r=R$, we have
\bbb
{\cal I}\big [  1  \big ]  = 2\m\, \area{d-1}  \left( {{2\m}\over\o} \right)\uu{{2\over 3}(d-2)} \ , 
\xxnn
\Delta H\lrm{edge} \ni   2\k\ll d\, \m\, \area{d-1} \left( {{2\m}\over\o} \right)\uu{{2\over 3}(d-2)} \ .
\een{EvaluationOfIntegratedIdentityEdgeTerm}

We have shown the 
integrated term ${\cal I}[1]$ is the only available
counterterm scaling as large as $\m\uu{+1}$ in $d=2$.  All other 
local edge terms of that size are ruled out by a combination
of conformal invariance and the ${\bf Y}$-dressing rule.
We conclude that the counterterm must be proportional to to 
${\cal I}$, with a possibly $d$-dependent numerical coefficient.


We emphasize that the coefficient $\k\ll d$ must really be
``numerical", rather than a ratio of scales $(\m / \o)\uu{d-{\rm dependent~exponent}}$, since a counterterm must be constructed
out of local observables and background couplings.  The only way one could possibly
get such a ratio as a local term would be to realize it as a term of the form
${\bf Y}\uu\a / {\bf Z}\uu\b$.  But edge operators containing fractional powers of ${\bf Z}$ 
are excluded by the dressing rule (Sec.~\ref{YDressingRule}).

Indeed, the coefficient $\k\ll d$ 
is fixed by the necessity of
cancelling the $(d-2)\uu{-1}$ term in the energy shift
$E\ls{c\ll 1}$ as calculated in \rr{SHCalculationOfEnergyShiftAsAnIntegralGeneralDimensionDIntermediateStage5}:
\bbb
(\Delta E)\ls {c\ll 1}\uprm{bulk} = -{{8\pi c\ll 1\m}\over{d-2}} - 8\pi c\ll 1\cc \m\cc  {\log}\cc\left(  {{2\m}\over\o}\right)
+ O\left[\m\uu 1 \cc (d-2)\uu 0\right] + O\left[(d-2)\uu 1\right]\ ,
\een{SHCalculationOfEnergyShiftAsAnIntegralGeneralDimensionDIntermediateStage5RECAP}
coming from the divergent integral of the $c\ll 1$ term in the bulk action for $d\leq2$.
In the usual way, to cancel the divergent term we must add ${\cal I}[1]$ with
a coefficient $\k\ll d \equiv {\k/({d-2})}$, determined by the condition
that the divergence cancel.  Using formula \rr{EvaluationOfIntegratedIdentityEdgeTerm} for the evaluation
of the droplet-edge term, we see that we need
\begin{eqnarray}
H\lrm{edge}\ni \k\ll d \cc {\cal I}[1]\cc \big |\ll{d\to 2} &=&  {\k\over{d-2}}\times {\cal I}[1]\cc \big |\ll{d\to 2}
\nonumber\\
\nonumber\\
&=&  {\k\over{d-2}}\times (2\m)\times \area{d-1}\cc \big |\ll{d\to 2} 
\nonumber\\
\nonumber\\
&=&  {{8\pi c\ll 1\m}\over{d-2}} \ ,
\end{eqnarray}
so we take $\k = 2\cc c\ll 1$,
which means
\bbb
\k\ll d  \equiv  {{2\cc c\ll 1}\over{d-2}}\ .
\eee
Thus, our edge-Hamiltonian counterterm, in conformal-edge minimal-subtraction, is
\bbb
(\Delta H)\lrm{edge} =   {{2\cc c\ll 1}\over{d-2}}\, {\cal I}[1]\ ,
\eee
or, in terms of the Lagrangian,
\bbb
(\Delta L)\lrm{edge} = - {{2\cc c\ll 1}\over{d-2}}\, {\cal I}[1]\ .
\eee
Written out explicitly, we have
\bbb
(\Delta L)\lrm{edge} = -  {{2\cc c\ll 1}\over{d-2}}\, m\uu{-2}\, \int \cc d\uu d \cc x \,
\d({\bf X})\, (m\sqd \cc {\bf Y})\uu{{d+4}\over 6}\ .
\een{NumericalCoefficientOfEdgeCountertermSpecified}

\subsection{Energy at order $c\ll 1\uu 1$ in $d=2$, with the counterterm included }
\label{DimRegPartC}

To calculate the contribution of the edge Lagrangian near $d=2$, including the finite term,
we have
\bbb
(\Delta E)\lrm{edge} = - (\Delta L)\lrm{edge}  = {{2\cc c\ll 1}\over{d-2}}\, \area{d-1}\,
m\uu{{{d-2}\over 3}}\,
 \cc R\uu{d-1}\, {\bf Y}\uu{{d+1}\over 6} \cc \bigg |\ll{r = R}\ .
\een{LocalEquationA}
Using eqns.~\rr{DefOfDropletRadius}, \rr{UnitSphereAreaValueRECAP0}, and \rr{BoldYInvariantClassicalAtDropletEdgeValue},
and expanding near $d=2$, we get
\bbb
(\Delta E)\lrm{edge} \to  {{8\pi\cc c\ll 1\cc\m}\over{d-2}} + {{16\pi \cc c\ll 1}\over 3} \,  \m\,
 {\log}\left( {\m\over\o}\right) + 4\pi\cc c\ll 1 \cc \g\lrm E \ .
\eenn
Ignoring the nonlogarithmic finite piece and restoring notation to reflect that we are working with 
the $O(c_1)$ sector of the theory, we have
\bbb
(\Delta E)\ls {c\ll 1}\uprm{edge} =  {{8\pi\cc c\ll 1\cc\m}\over{d-2}} + {{16\pi \cc c\ll 1}\over 3} \cc\m\cc
 {\log}\left({\m\over\o}\right) + O\left[ \m\uu 1 \cc (d-2)\uu 0 \right]\ .
\eee
Adding the counterterm 
contribution to the bulk contribution
\rr{SHCalculationOfEnergyShiftAsAnIntegralGeneralDimensionDIntermediateStage5}, 
the total is
\bbb
(\Delta E)\ls {c\ll 1}\uprm{total} = (\Delta E)\ls {c\ll 1}\uprm{bulk} +
(\Delta E)\ls {c\ll 1}\uprm{edge} = - {{8\pi\cc c\ll 1}\over 3} \cc \m
\cc {\log}\left(  {\m\over\o} \right)
+ O\left[\m\uu 1 \cc (d-2)\uu 0 \right] \ ,
\een{DimRegSuccessfulRecalculationResult}
where the latter part is finite, scheme-dependent, and nonlogarithmic.  The $\m\uu 1$ term can be absorbed into the finite part of the
coefficient of the counterterm $\d({\bf X}){\bf Y}$.
Note, of course, that the coefficient of the $\m\cc{\log}(\m / \o)$ term
agrees with the value \rr{FinalAnswerOrderC1EnergyShiftInSharpCutoffRenormalized} computed in a conformally-invariant
sharp-cutoff regulator in Sec.~\ref{ConformalSharpCutoff}, 
which also agrees with the value computed with essentially the same type of regulator in \cite{Kravec:2018qnu}.
Written in terms of $Q$, using the relations \rr{SimplyExpressedRelationshipBetweenMuAndQInTheHarmonicTrapRECAPFirstAppearanceRestrictedToTwoDimensions},\rr{SimplyExpressedInverseRelationshipBetweenMuAndQInTheHarmonicTrapRECAPFirstAppearanceRestrictedToTwoDimensions},
we have
\bbb
(\Delta E)\ls {c\ll 1}\uprm{total} = - {{4\pi\cc c\ll 1}\over 3} \cc \xi\uu{\hh} Q\uu{\hh} \o
\cc {\log}\left(  \xi Q \right)
+ O\left(\o \xi\uu\hh Q\uu\hh  \right)\ .
\een{DimRegResultInXi}

There is some ambiguity in the extension of
the definition of $\xi$ in ref \cite{Son:2005rv} to general dimension $d$, that translates into an order $(d-2)\uu 0\cc \m\uu 1$ term in the energy, in $d=2$.  This term can be re-absorbed into the $(d-2)\uu 0$ piece of the coefficient of the counterterm.  We give
the details of the scheme-dependence in section \ref{ConventionsForXiDefinitionSummary} of the Appendix,
and discuss its consequences for the scheme-dependence
of the finite piece of the coefficient of the $\m\uu{+1}$ edge counterterm in $d=2$,
in Sec.~\ref{ConventionSensitivityOfCounterterms}. 

\subsection{Contributions of other subleading operators with positive $Q$-scaling in $d=2$}

So far we have analyzed in detail the effect of a bulk subleading operator \rr{NLOTerm1}, \rr{YDef} 
and a boundary counterterm \rr{DressedIdentityEdgeOp} on the ground state energy in $d$ dimensions,
particularly in the case $d=2$, in which the bulk contribution is logarithmically divergent.  
We did this to motivate a systematic analysis of boundary counterterms and their classical contributions 
to the vacuum energy.  We have not yet analyzed the effects of other subleading operators
that contribute to the ground state energy with nonnegative powers of $Q$, because the analysis of those terms proceeds very similarly to the cases we have already considered.  For completeness, we can briefly present the results.

\subsubsection{The bulk term ${\cal L}\ll{c\ll 2}$}\label{EffectOfTheOtherBulkTermInTwoDimensionsAtLeast}
First we consider the term ${\cal L}\lrm{c2} $ written in \rr{NLOTerm2}, which we recap here:
\bbb
{\cal L}\lrm{c2} \equiv - c_2\, d\sqd\cc m\uu{\hh(d-2)}\cc {\bf X}\uu{{{d\over 2}} - 1}\cc{\bf Z}\ ,
\xxn{NLOTerm2Recap}
{\bf Z} \equiv   \vec{\pp}\sqd\cc A\ll 0 - {1 \over{ d^2\cc m}} (\vec{\pp}\sqd\cc \chi)^2 \ .
\een{ZDefRecap}
In contrast to the contribution of the $c\ll 1$ term, 
the contribution of the $c\ll 2$ term is convergent 
in $d=2$.\footnote{The $c\ll 2$ contribution to the energy in the harmonic potential is also convergent
in $d=3$ as well; the result is given in Eqn.~(9) of \cite{Son:2005rv}.} 
The classical value of ${\bf Z}$ in the harmonic potential \rr{HarmonicTrappingPotentialDef} is 
\bbb
{\bf Z}\lrm{classical} = m\o\sqd d\ ,
\eee
and so in $d=2$ we have 
\bbb
{\bf Z} = 2 m \o\sqd\ .
\een{ZValueTwoDims}
In $d=2$ there is no ${\bf X}$-dressing at all of the ${\cal L}\ll{c\ll 2}$ term, and so
\bbb
{\cal L}\lrm{c2} = - 4 c\ll 2 {\bf Z} = - 8 c\ll 2 m \o\sqd\ .
\eee
Integrated over the extent of the droplet, this is just the droplet area times $-2 c\ll 2 m \o\sqd$:
\bbb
(\Delta E)\ls{c\ll 2} = -(\Delta L)\ls{c\ll 2} =  8\pi c\ll 2 m \o\sqd R \sqd = 16\pi c\ll 2 \m\ .
\eee
This term is obviously finite, and contributes parametrically at the same scale 
as the boundary counterterm $\d({\bf X}) {\bf Y}$ in two dimensions. 

\subsubsection{The edge term $\co\ll{b\ll 2}\equiv [m\uu{-1} {\bf Z}]\lrm{ edge} $}\label{ContributionOfDressedZEdgeOperator}

The operator ${\bf Z}$ also participates in the edge term 
\bbb
\co\ll{b\ll 2} \equiv [m\uu{-1} {\bf Z}]\lrm{ edge} 
\equiv m\uu{-1}\cc \d({\bf X}) (m\sqd {\bf Y})\uu{{1\over 3}} {\bf Z}\ ,
\eee
which is nonzero classically, and whose integrated
contribution scales as $\m\uu{{1\over 3}}$ in $d=2$, by formulas \rr{GammaTotalGeneralRepresentation}, \rr{GammaTotalFormula}.
Concretely, by \rr{ZValueTwoDims}, \rr{BoldYInvariantClassicalAtDropletEdgeValue}, and \rr{DefOfDropletRadius}, we have the classical value
\begin{eqnarray}
\langle \co\ll{b\ll 1} \rangle &=& \langle  [m\uu{-1} {\bf Z}]\lrm{ edge}  \rangle = 
m\uu{-{1\over 3}}\cc \langle {\bf Z}\cc {\bf Y}\uu{- {1\over 6}}\rangle \cc \d(|x| - R)\ ,
\nonumber\\
\nonumber\\
\langle {\cal I}[m\uu{-1}{\bf Z}] \rangle &\equiv& \int\cc d\sqd x  \cc \langle \co\ll{b\ll 1} \rangle =  
2\pi R \, m\uu{-{1\over 3}}\cc \langle {\bf Z}\cc {\bf Y}\uu{- {1\over 6}} \rangle
\nonumber\\
\nonumber\\
&=& 4\pi\,(2\o\sqd\m)\uu{{1\over 3}}\ .
\end{eqnarray}
This term does not arise as a UV-divergent
counterterm in either of the conformal regulators we have considered in $d=2$ (either
the conformal sharp cutoff or conformal dimensional regularization).  It could arise in principle as a divergent counterterm in some other conformal
cutoff, but we do not know of one.

In any complete NRCFT realizing the conformal EFT,
including the unitary fermi gas, one expects all possible effective terms
to arise, and so one expects the edge operator $[m\uu {-1} {\bf Z}]\lrm{edge}$ to appear with a finite coefficient.  It would be interesting
to learn the coefficient of this edge operator in the unitary fermi gas,
by any experimental, numerical, or theoretical methods available.  Possible theoretical tools 
might include a nonrelativistic analog\footnote{See {\it e.g.} \cite{Goldberger:2014hca} for comments on the subject} of the 
large-charge conformal bootstrap \cite{Jafferis:2017zna}, or a nonrelativistic analog of the powerful
large-charge double-scaling techniques recently invented for the study of nonsupersymmetric \cite{Arias-Tamargo:2019xld, Arias-Tamargo:2019kfr, Arias-Tamargo:2020fow, Badel:2019oxl, Badel:2019khk} and supersymmetric \cite{Bourget:2018obm, Bourget:2018fhe, Grassi:2019txd} relativistic conformal field theories.

\subsection{Upper bound on quantum $\m$-scalings of contributions of edge operators to the ground state energy}\label{CrudeUpperBoundQuantumScalings}
We conclude this section with a note on quantum corrections, which requires giving an estimate 
of the quantum mechanical $\m$-scalings of operators with vanishing expectation value in the classical solution.  
The estimate entails an analysis of the regularization and renormalization of the quantum fluctuations of the $\chi$ field in the harmonic trap.
We defer a detailed analysis to later work \cite{toAppearSHIS}, but in this section we will give a loose upper bound that is
sufficient to ensure that only a finite number of operators can contribute quantum mechanically to an observable at or above a given order
in $\m$, in any given spatial dimension $d$.

As a particular application, we will show that in $d=2$ there are no tree or loop graphs contributing at order $Q\uu 0$, other than the one-loop vacuum bubble representing the Casimir energy.
This implies that the order $Q\uu 0$ term in the dimension of the lowest operator with particle number $Q$ in two spatial dimensions
is universal and calculable.

\subsubsection{UV cutoff and scaling of the propagator}

We now briefly consider the quantum effects of operators with vanishing expectation value at the classical level in the 
ground-state solution in the harmonic potential. 
The bare quantum expectation value of 
an operator is cutoff-dependent, and this cutoff-dependence must be
considered carefully for these operators, since the UV-divergent
quantum contribution is the leading effect controlling the $\m$-scaling.

In considering these contributions it is important to recall the hierarchical 
separation \rr{DoubleHierarchyForPerturbativeRenormalization} between the Wilsonian cutoff and the UV
scale.  In the limit $Q\to\infty$, the cutoff on energy and/or momentum is taken to be parametrically lower than 
the UV scale set by $\m$.  So when counting the quantum contributions
of operators, time and/or spatial derivatives never contribute positive powers of $\m$, even at the quantum level 
in UV-divergent diagrams.

To obtain an estimate, we must bound the scaling of the $\chi$ propagator.  
The Gaussian terms in the Lagrangian density go as $\m\uu{{d\over 2} - 1}\dot{\chi}\sqd$ and
$\m\uu{{d\over 2}}(\pp\ll x \chi)\sqd$.  
At energies of order $\o$ and momenta of order $R\uu{-2} \sim \o\sqd / \m$, the $\chi$ propagator then goes as 
\bbb
\langle \hat{\chi}(p,E) \hat{\chi}(p,-E) \rangle \sim \m\uu{1 - {d\over 2}} \o\uu{-2}\ , 
\eee
while $E \sim \o$ and $p\sim R\uu{-1} \sim \o \sqrt{m/\m}$.
The anisotropy of the propagator and the singularity near the boundary 
complicate the analysis, and we postpone a detailed treatment to later work \cite{toAppearSHIS}.  For now,
we will compute our estimate by imposing an energy cutoff $\L$ that is independent of $\m$, and a 
momentum cutoff $p\lrm{max} = m\uu\hh \L / \sqrt{\m}$ that goes to zero as the inverse
square root of the chemical potential.  This cutoff suffices for an analysis of the quantum 
contributions to the vacuum energy in the harmonic potential, since the infrared momentum
scale is $R\uu{-1} \sim m\uu\hh \o / \sqrt{\m}$, so we still have $\sqrt{m\m} \muchgreaterthan p\lrm{max} \muchgreaterthan R\uu{-1}$, 
so long as we take $\m \muchgreaterthan \L \muchgreaterthan \o$:
\bbb
p < p\lrm {max} \equiv  \L \sqrt{m / \m} \ , \qquad E < \L \ , \qquad \L \muchlessthan \m\ .
\eee
With this cutoff, the quantum scaling of a fluctuation $\hat{\chi} \equiv \chi - \langle\chi\rangle$ is given by the square root of the propagator,
\bbb
\hat{\chi} \sim \m\uu {{2-d}\over 4}\ .
\eee

\subsubsection{A crude bound on the quantum $\m$-scaling of general edge operators}

Thus, multi-derivatives of $\hat{\chi}$ scale as
\bbb
\co\ll{n\ll x , n\ll t}\equiv \pp\ll x\uu {n\ll x} \pp\ll t\uu {n\ll t} \hat{\chi} \sim \m\uu{-{{d-2}\over 4} - {{n\ll x}\over 2}}\ .
\eee
We then have that
\begin{eqnarray}
\b(\co\ll{n\ll x , n\ll t}) &=& -{{d-2}\over 4} - {{n\ll x}\over 2}\ ,
\nonumber \\
\nonumber \\
\D(\co\ll{n\ll x , n\ll t}) &= &n\ll x + 2n\ll t\ , 
\nonumber \\
\nonumber \\
\g\lrm{term} (\co\ll{n\ll x , n\ll t}) &=& -{{d-2}\over 4} - {{n\ll x}\over 2} - {{n\ll x}\over 6} - {{n\ll t}\over 3} = - {{3d-6+8 n\ll x + 4 n\ll t}\over{12}} \ ,
\end{eqnarray}
for each multiderivative of $\chi$, and
\bbb
\g\lrm{term}(\co\lrm{undressed}) = -{1\over{12}}\cc \sum\ll a 3d-6+8 n\ll x\upp a + 4n\ll t\upp a\ ,
\eee
for a general monomial in differentiated fields.

We can improve the bound further with some simple considerations.  For each $\chi$, there must be at least one $\pp\ll x$ or one $\pp\ll t$, by
global charge conservation.  First, consider the case $n\ll x = 0$ and $n\ll t = 1$, that is, the term $\dot{\hat{\chi}} \equiv \dot{\chi} - \m$.   
The term $\dot{\chi}$ must be completed to ${\bf X}$ by gauge invariance and conformal invariance,
so we must consider the quantum $\m$-scaling of $\hat{{\bf X}} \equiv {\bf X} - \langle{\bf X}\rangle$.
But ${\bf X}$ vanishes as an operator\footnote{It vanishes identically, not just as an expectation value.} at the edge. 
Multiplying by the factor $\d({\bf X})$ in the dressing kills the operator: ${\bf X}\, \d({\bf X})  = 0$.
So there is no term with $n\ll t = 1$ and $n\ll x = 0$.

Therefore, we can assume that $2n\ll x + n\ll t \geq 2$ for a nonvanishing, globally symmetric egde term.  This means
\bbb
3d-6+8 n\ll x\upp a + 4n\ll t\upp a \geq 3d+2\ ,
\eee
and so
\bbb
\g\lrm{term}(\co\lrm{undressed}) \leq -{{3d+2}\over{12}}\cc n\ll\chi\ ,
\eee
giving an upper bound on the total $\m$-scaling of such a term after dressing and integration:
\bbb
\g\lrm{total} \leq {{2d-1}\over 3}-{{3d+2}\over{12}}\cc n\ll\chi = {1\over{12}}\big [ 8d-4 - (3d+2)n\ll\chi ]\ .
\een{TotalQuantumMusScalingOfDressedIntegratedTerm}
The estimate \rr{TotalQuantumMusScalingOfDressedIntegratedTerm}
is weaker than one might like: It allows a growing number of operators contributing with positive $Q$-scaling to the large-charge 
vacuum energy in the harmonic potential.  The maximum number of $\chi$-fluctuations in an operator 
with positive $\g\lrm{total}$ grows with spatial dimension as $n\ll\chi\uprm{max} \sim {8\over 3} (d-2)$.
With some further refinement, the bound can be strengthened and the number of contributing 
operators reduced \cite{toAppearSHIS}, but the simple bound 
\rr{TotalQuantumMusScalingOfDressedIntegratedTerm} suffices to make the classification of 
contributing edge operators tractable in low spatial dimension.  In particular, for $d=2$ we can show
that there are no edge terms making quantum contributions to the vacuum energy at order $Q\uu 0$ or greater.

\subsubsection{No quantum corrections to the vacuum energy with nonnegative $\m$-scaling in $d=2$}

For $d=2$, the only edge terms that could possibly contribute with a positive $\m$-scaling 
are $\dot{{\bf X}}$ and $\gg\ll x\chi$, both of which have integrated dressed $\m$-scaling $\leq {1\over 3}$, which is order $\m\uu{-{8\over 3}}$
relative to the leading term $\m\uu 3$.  Both of these undressed operators vanish in the 
classical ground state solution, and can contribute to the vacuum energy in the harmonic potential only
through their quantum effects.  Since the one-point functions of these terms vanish 
at tree level, the only possible contributions can come from tree-level contributions with more than one insertion, or from loop corrections.

Connected contributions with more than one insertion are too small 
to contribute with positive $\m$-scaling.  Each insertion of a (dressed and integrated) $\dot{{\bf X}}$ term or $\gg\ll x\chi$ 
term in the path integral would lower the $\m$-scaling by ${8\over 3}$ relative to the
leading-order energy $\m\uu 3$.   At least 
two insertions would needed to get a nonvanishing contribution, which would bring down the total $\m$-scaling of the 
two-insertion effect to $- {7\over 3}$.  Thus, in $d=2$ we conclude that tree graphs involving vevless operators can only
contribute terms with negative $\m$-scaling to the vacuum energy.

Loop corrections to the contribution with one insertion are also too small to contribute. 
At tree level, the largest edge operator is the dressed identity, whose integral ${\cal I}[1]$ enters at order $\m\uu{{{2d-1}\over 3}}$.
The action in $d$ dimensions scales as $\m\uu {1+d}$, so the loop-counting parameter is
$\m\uu{-(d+1)} $.  
Thus, a one-loop correction to the contribution of ${\cal I}[1]$ can be no larger than $\m\uu{-{{d+4}\over 3}}$.
We conclude that contributions with even one loop and one edge operator all have negative $Q$-scaling in any dimension, including $d=2$.

\section{Conclusions}\label{conclusions}
In this paper we have done the following:

\bi
\item{We have given a concrete representation for operators localized at the edge of the region of support of the particle droplet
in a nonrelativistic CFT, where the density falls to zero.}
\item{We have bounded the classical and quantum contributions of higher-derivative operators at 
the droplet edge, reducing the set of operators contributing at any given order in the large-$Q$ expansion to a finite set.}
\item{We have used our classification of edge operators to demonstrate the scheme-independence of the
$Q\uu{\hh} {\log}(Q)$ term in the operator dimension in
two spatial dimensions.}
\item{We have uncovered a second edge operator contributing 
to the lowest charged operator dimension with a positive $Q$-scaling; this operator goes as $\m\uu{{1\over 3}}$ in $d=2$ and
as $\m\uu{+1}$ in $d=3$.  This coefficient of this edge operator is scheme-independent in the sense that it does not renormalize a bulk UV divergence.  The appearance of
edge terms with third-integer powers of the chemical potential is an interesting and unexpected outcome of the operator analysis, reminiscent 
of the appearance of quarter-integer powers of the angular
momentum for effective string theory at large spin \cite{Hellerman:2016hnf} \cite{Sever:2017ylk}.}
\ei
While a 
detailed analysis of the quantum effects of
conformal edge operators is postponed to future work \cite{toAppearSHIS}, a coarse bound on the size of quantum effects shows that no term contributes
classically or quantum-mechanically at order
$Q\uu 0$ or larger in $d=2$.  This allows a straightforward
calculation of the one-loop vacuum energy 
in the harmonic trap in $d=2$, with the result:
\bbb
E\big |\ll{O(Q\uu 0)} = - 0.29416\cc\o\ .
\een{UnderivedValueCasimirEnergyTwoDimensionsNRCFT}
The result \rr{UnderivedValueCasimirEnergyTwoDimensionsNRCFT} is a renormalized Casimir-type 
energy obtained by a renormalized Coleman-Weinberg sum over the energy 
spectrum \rr{PhononEnergySpectrum}.  Due to the absence of an 
order $Q\uu 0$ counterterm in $d=2$, the renormalized value of this term is universal and unambigious.

More broadly, we have derived
an asymptotic expansion of the ground state energy of the $Q$-particle
state in a harmonic potential with trapping frequency $\o$ in $d=2$, with the structure
\begin{eqnarray}
E &=&  E\ll{{3\over 2}} + E\ll{\hh} + E\ll{{1\over 6}}
+ E\ll 0 + ({\rm negative~}Q{\rm-scalings})\ ,
\nonumber\\
\nonumber\\
E\ll{{3\over 2}} &\equiv&
{2\over 3}\cc \xi\uu{\hh} Q\uu{{3\over 2}} \cc\o \ ,
\nonumber\\
\nonumber\\
E\ll{{1\over 2}} &\equiv & - {{4\pi\cc c\ll 1}\over 3} \cc \xi\uu{\hh} Q\uu{\hh} 
\cc {\log}\left(  \xi Q \right) \cc \o
+ \a\ll {\hh} Q\uu{\hh}\o 
\nonumber\\
\nonumber\\
E\ll{{1\over 6}} &\equiv& \a\ll{{1\over 6}}\cc Q\uu{{1\over 6}}\cc \o\ ,
\nonumber\\
\nonumber\\
E\ll 0 &\equiv& - 0.29416\cc \o \ .
\end{eqnarray}
The leading term $E\ll{{3\over 2}}$ is determined solely by the
leading term $c\ll 0 m  {\bf X}\sqd$ in the action.  The logarithmic
part of the $E\ll{\hh}$ term comes solely from the ${\cal L}\ll{c\ll 1}$
term in the EFT, while the nonlogarithmic term is scheme-dependent,
depending on the details of
the cutoff procedure, on the coefficients of the $\d({\bf X}) {\bf Y}$
edge term, and also on the coefficient of the ${\bf Z}$ bulk term.  The
$E\ll{{1\over 6}}$ term is scheme-independent and
depends only on the (finite) coefficient of the edge term $\d({\bf X}){\bf Y}\uu{-{1\over 3}} {\bf Z}$.  This term is gauge-invariant, conformally invariant,
and by naturalness presumably appears with a nonzero coefficient in the edge Hamiltonian of any UV completion of the Son-Wingate EFT,
including the interacting NRCFT describing the unitary fermi gas at
quantum criticality.  To our knowledge, this term and its higher-dimensional
analogs have not been discussed in any analysis of the unitary fermi gas,
and it would be interesting to constrain its coefficient by any available means.

The $E\ll 0$ coefficient, the explicit details of whose calculation we defer to
future work \cite{toAppearSHIS}, is universal, not just among all
UV completions of the critical unitary fermi gas, but among any other $2+1$-dimensional nonrelativistic CFT that lies in the
same ``large-charge universality class" as that of the critical unitary
fermi gas.\footnote{See for example Sec.~7 of \cite{Nishida:2010tm} 
and references therein for a discussion of other NRCFTs, some of which may lie in the same (or a similar) universality class 
as the unitary fermi gas at large charge.} Such distinct NRCFTs would have
other values for the coefficients $\xi, c\ll 1, \a\ll {\hh}, \a\ll{{1\over 6}}$, 
and so on, but should exhibit the same structure of the asymptotic expansion,
the same leading-order excitation spectrum, and the same value of the
universal coefficient of the $E\ll 0$ term.  Though we are not aware of any specific examples, one 
possible application of the present work may be to suggest possible NRCFTs in
the same large-charge universality class, which could be constructed using the large-charge expansion 
as a clue to the structure of the full theory.

Of course, by the NRCFT state-operator correspondence, the terms above
correspond to the asymptotic expansion of the scaling dimensions of the 
lowest charged operators in the NRCFT with particle number $Q$:
\begin{eqnarray}
\D(Q) 
&=& \D\ll{{3\over 2}}  + \D\ll{\hh} + \D\ll{{1\over 6}}
+ \D\ll 0 + ({\rm negative~}Q-{\rm scalings})\ ,
\nonumber \\
\nonumber \\
\D\ll{{3\over 2}} &\equiv&
{2\over 3}\cc \xi\uu{\hh} Q\uu{{3\over 2}}  \ ,
\nonumber \\
\nonumber \\
\D\ll{{1\over 2}} &\equiv& - {{4\pi\cc c\ll 1}\over 3} \cc \xi\uu{\hh} Q\uu{\hh}\cc {\log}\left(  \xi Q \right) 
+ \a\ll {\hh} Q\uu{\hh}
\nonumber \\
\nonumber \\
\D\ll{{1\over 6}} &\equiv& \a\ll{{1\over 6}}\cc Q\uu{{1\over 6}}\ ,
\nonumber \\
\nonumber \\
\D\ll 0 &\equiv& - 0.29416\ .
\end{eqnarray}
As a final advertisement, we will analyze the quantum effects of edge counterterms
and give details of the computation of the
universal $Q\uu 0$ term \rr{UnderivedValueCasimirEnergyTwoDimensionsNRCFT}, in the future \cite{toAppearSHIS}.

\section*{Note}

While this paper was in preparation we learned of another forthcoming work with related results \cite{ToAppearReffertEtAl}.

\begin{appendix}

\section{Conventions}\label{ConventionAppendix}

For convenience, we include here a dictionary of translations among conventions in the recent literature.

\subsection{Summary of relationships among the variously-defined
$\xi$-coefficients and the $c\ll 0$ coefficient in the recent literature }\label{ConventionsForXiDefinitionSummary}

We define $\xi$ in this paper the same was as defined in \cite{Nishida:2010tm}.
There, $\xi$ is defined as the ratio of the unitary Fermi gas ground state
energy density at a given fermion density, to the energy density
of the free Fermi gas at the same fermion density:
\bbb
\xi \equiv {{{\cal H}\lrm{interacting}(\r)}\over{{\cal H}\lrm{free}(\r)}}\ , 
\een{UnderSpecifiedExpression}
where ${\cal H}$ is the ground-state energy density in the infinite-volume limit at fermion density $\r$, 
with vanishing background potential, in $d$
dimensions.  However, the denominator in expression \rr{UnderSpecifiedExpression}
is not well-defined as an arbitrary function of spatial dimension $d$
without further information.  Specifically, as noted in the text above, the energy density at a given
fermion density depends on the number $a\ll s$ of spin-and-species states.
Ref.~\cite{Nishida:2010tm}'s convention is to define {\it the free fermi gas} as a
system of fermions with two identical spinless fermion species,
for all spatial dimensions $d$.  This convention is not stated explicitly, but can be
inferred from the formula for the (free) fermi momentum {$k\lrm F$} in terms of
the free fermion density, which \cite{Nishida:2010tm} denotes by {$n$}, equivalent to our $\r$. 
The formula, given below eqn.~{(18)} of \cite{Nishida:2010tm}, states
\bbb
k\lrm F = [ 2\uu{d-1} \pi\uu{d/2} \G({d\over 2} + 1)\cc n]\uu{{1\over d}} \ ,\qquad (a_s = 2\ {\rm via\ Nishida-Son})\ ,
\een{NishidaSonDefinitionOfFermiMomentumTwoScalarSpeciesImplicitlyAssumed}
in general spatial dimension $d$. The relationship between the ground state energy density and the Fermi momentum depends of course on
the number $a\ll s$ of free fermion states per momentum level; in terms of $a\ll s$ the general relationship is
\bbb
k\lrm F = [ a\ll s\uu{-1}\cc 2\uu d \pi\uu{d/2} \G({d\over 2} + 1) \cc \r]\uu{{1\over d}}\ .
\een{FermiMomentumGeneralNumberOfScalarSpecies}
So, the more precise characterization of $\xi$ with \cite{Nishida:2010tm}'s implicit definition is to express 
\bbb
   \xi \equiv  {{{\cal H}\lrm{interacting}(\r)}\over{{\cal H}\lrm{free}(\r)\cc \big |\ll{a\ll s = 2}}}
 \ .
\eee
Given this definition of $ \xi$, the relationship to the coefficient $c\ll 0$ in the effective Lagrangian is
\bbb
\xi 
= (2\pi)\uu{-1} \, \left( \cc      { \frac{1}{2} {\G\left({d\over 2} + 2 \right)}}  \, c\ll 0 \cc \right)\uu{-{2\over d}}\ ,
\xxn{XiInTermsOfNishidaSonC0SummaryInAppendix}
c_0 = {2\over{ \G\left({d\over 2} + 2\right)}}\,\, (2\pi)\uu{-{d\over 2}}\,\, \xi\uu{-{d\over 2}}\ .
\een{C0InTermsOfNishidaSonXiSummaryInAppendix}

The convention for the definition of the Bertsch coefficient $\xi$ in \cite{Kravec:2018qnu} is slightly different.
Denoting the coefficient appearing there by $\xi\ls{\rm KP}$, we have
\bbb
\xi\ls{\rm KP} = \left( \cc \hh\cc \G(d+1) \cc \right)\uu{{1\over d}} \, \xi\uu{\hh}
\xxn{NishidaSonXiInTermsOfKravecPalXi}
\xi
=\left( \cc {2\over{ \G(d+1)}} \cc \right)\uu{{2\over d}} \,
\xi\ls{\rm KP}\sqd\ ,
\een{KravecPalXiInTermsOfNishidaSonXi}
so that the relationship between the $c\ll 0$ coefficient and the coefficient ${\xi}\ls{\rm KP} $ of
 \cite{Kravec:2018qnu} is given by
\bbb
{\xi}\ls{\rm KP}  =
\bigg [ \cc c\ll 0 \,
{{ (2\pi)\uu{{d\over 2}}\times \G\big(2 + {d\over 2} \big )}\over{\G\big ( 1 +  d \big )}}
\cc\bigg ]\uu{-{1\over d}} \ ,
\xxn{EquationGivingKravecPalsXiInTermsOfC0RECAP2Summary}
c\ll 0 =\bigg [ \cc
{{ (2\pi)\uu{-{d\over 2}}\times \G\big(  1 +  d \big )}\over{\G\big ( 2 + {d\over 2} \big )}}
\cc\bigg ] \, \big ( \cc {\xi}\ls{\rm KP} \cc \big )\uu{-d}\ ,
\een{EquationGivingC0InTermsOfKravecPalsXiRECAP3Summary}

\subsection{Conventions for NLO coefficients}\label{NLOCoefficientConventionsAppendix}

In eqns.~\rr{NLOStructureHere}, \rr{NLOTerm1}, and \rr{NLOTerm2}, the NLO bulk Lagrangian is given by
\bbb
{\cal L} = m\uu{\hh(d-2)}\cc c\ll 1\cc {\bf X}\uu{{d\over 2} - 2}\cc{\bf Y} -d\sqd\cc m\uu{\hh(d-2)}\cc c\ll 2\cc {\bf X}\uu{{{d\over 2}} - 1}\cc{\bf Z}\ ,
\een{BulkLagrangianNLOStructureRecap}
with the terms normalized in \rr{YDef}, \rr{ZDef} as
\bbb
{\bf Y} \equiv (\vec{\pp} {\bf X})\sqd
\xxn{YDefRecapAgain}
 {\bf Z} \equiv  [ \vec{\pp}\sqd\cc A\ll 0 - {1 \over{ d^2\cc m}} (\vec{\pp}\sqd\cc \chi)^2 ]\ .
\een{ZDefRecap}
These are the same conventions used in \cite{Son:2005rv} (SW), albeit restricted to $d=3$:
\bbb
c\ll {1,~{\rm \hbox{SW}}} = c\ll {1,~{\rm here}} \big |\ll{d=3}\  , \qquad
c\ll {2,~{\rm \hbox{SW}}} = c\ll {2,~{\rm here}} \big |\ll{d=3}
\ .
\eee
Our conventions are also the same as those of \cite{Favrod:2018xov}, except that ref.~\cite{Favrod:2018xov}
sets the background gauge potential to zero:
\bbb
{\cal L}\ll{1,~{\rm Favrod\ et\ al.}} =  {\cal L}\ll {1,~{\rm here}} \ , \qquad
{\cal L}\ll{2,~{\rm Favrod\ et\ al.}} =  {\cal L}\ll {2,~{\rm here}}\big |\ll{A\ll 0 \to 0} \ ,
\eee
with
\bbb
c\ll {1,~{\rm  Favrod\ et\ al.}} = c\ll {1,~{\rm here}} \ ,
\llsk\llsk
c\ll {2,~{\rm  Favrod\ et\ al.}} = c\ll {2,~{\rm here}} \ .
\eee
Also note that the derived quantity $d\ll 0\sqd$, defined below eqn.~{(3.5)} of \cite{Favrod:2018xov}, is given by
\bbb
d\ll 0\sqd ={1\over 4}\times \G\left({d\over 2} \right)\,\, (2\pi)\uu{+{d\over 2}}\,\, \xi\uu{{d\over 2}}
\een{ValueOfLittleDNotSquaredFromFavrodEtAl}
when written in terms of $\xi$.

\subsection{Comparison of names for coefficients and local quantities, making explicit the factors of $m$ and $\hbar$}\label{ConventionComparisonLists}

To proceed further, we note some differences in conventions for dimensional analysis, among various relevant
works in the literature:
\bi
\item{Ref.~\cite{Kravec:2018qnu} sets $m\to 1$ but \cite{Son:2005rv,Favrod:2018xov} do not, and neither do we;}
\item{For better or worse, we, \cite{Son:2005rv}, and \cite{Kravec:2018qnu} all set $\hbar \to 1$, though \cite{Favrod:2018xov} does not.}
\ei
With these in mind, the reader may apply:
\bbb
\th\ll{\left [ \SWMacroA\right ]} =
\th\ll{\left [ \FORMacro\right ]} =
 \chi\ll{\left [ \KPMacroA \right ]} = \chi\ll{\left[\SHISHereMacro\right]}\ ,
 \xxx
{\bf X} \ll{\left [ \SWMacroA\right ]}  = {\bf X}\ll{\left [ \KPMacroA \right ]} = 
{\cal U}\ll{\left [ \FORMacro \right ]} 
 = {\bf X}\ll{\left[\SHISHereMacro\right]}\ ,
\xxx
n\ll{\left [ \SWMacroA\right ]} = n\ll{\left [ \KPMacroA \right ]} 
= \r\ll{\left[\SHISHereMacro\right]}
= {{\d{\cal L}}\over{\d\dot{\chi}}} = \hbar\cc\r\ll{\left [ \FORMacro \right ]} \ ,
\xxx
m \ll{\left [ \SWMacroA\right ]} = m\ll{\left[\SHISHereMacro\right]} = m\ll{\left [ \FORMacro \right ]}
= 1\ll{\left [ \KPMacroA \right ]}\ .
\eee

At the level of the leading-order Lagrangian, and in utterly explicit detail:
\begin{eqnarray}
P({\bf X})  \ll{\left [ \SWMacroA\right ]} &= & {\cal L} \ll{\left [ \SWMacroA\right ]}  \qquad {\rm(by\ eqn.~(62)\ of\ Son-Wingate)  }
\nonumber\\
\nonumber\\
 &= & {\cal L}\ll{\left[\SHISHereMacro\right]} \qquad {\rm(because\ the\ action \ is\ the\ action)}
 \nonumber\\
 \nonumber\\
  & =& {\cal L} \ll{\left [ \KPMacroA\right ]}
  \nonumber\\
 \nonumber\\
  &=& {1\over\hbar}\cc {\cal L}\ll{\left [ \FORMacro \right ]} \qquad {\rm(because\ Favrod\ et\ al.\ keep\ \hbar)}
  \nonumber\\
  \nonumber\\
  &=& \bigg ( \cc c\ll 0\cc {\bf X}\uu{1 + {d\over 2}} \cc \bigg )\ll{\left [ \KPMacroA\right ]} \qquad  {\rm(they\ set\ m\to 1)}
  \nonumber\\
  \nonumber\\
  &=&   \bigg ( \cc m\uu{d\over 2}\cc c\ll 0\cc {\bf X}\uu{1 + {d\over 2}} \cc \bigg )\ll{\left [ \SWMacroA  /  {{\rm Nishida-}\atop{\rm Son}} \right ]}  
  \nonumber\\
  \nonumber\\
    &=&  \bigg ( \cc \hbar\uu{-{d\over 2}}\cc
   m\uu{d\over 2}\cc c\ll 0\cc {\cal U}\uu{1 + {d\over 2}} \cc \bigg )\ll{\left [ \FORMacro \right ]}  \qquad  {\rm(they\ keep\ \hbar\ explicit)} 
   \nonumber\\
   \nonumber\\
   &=&  \bigg ( \cc m\uu{+{d\over 2}}\cc
     c\ll 0\cc {\bf X}\uu{1 + {d\over 2}} \cc \bigg )\ll{\left [ \SHISHereMacro\right ]}  \qquad  {\rm(we\ do\ not\ keep\ \hbar\ explicit)}\ .
\end{eqnarray}
So the dictionary of $c\ll 0$-coefficients is
\bbb
c\ll 0\uu {\left [ \KPMacroA\right ]} = 
m\uu{{d\over 2}}\cc c\ll 0\uu {\left [ \SWMacroA\right ]}
= 
m\uu{{d\over 2}}\cc c\ll 0\uu {\left [ \SHISHereMacro\right ]}
 =  \hbar\uu{-{d\over 2}}\cc
   m\uu{{d\over 2}}\cc c\ll 0\uu{\left [ \FORMacro \right ]} \ ,
   \xxn{ConventionComparisonOfC0Coefficients}
   c\ll 0\uu {\left [ \SHISHereMacro\right ]}
 = c\ll 0\uu {\left [ \SWMacroA\right ]}
=  \hbar\uu{-{d\over 2}}\cc
   c\ll 0\uu{\left [ \FORMacro \right ]} 
   = m\uu{-{d\over 2}}\cc c\ll 0\uu {\left [ \KPMacroA\right ]} \ ,
   \xxx
    c\ll 0\uu{\left [ \FORMacro \right ]} 
   = \hbar\uu{{d\over 2}}\cc
    c\ll 0\uu {\left [ \SHISHereMacro\right ]}
 = \hbar\uu{{d\over 2}}\cc c\ll 0\uu {\left [ \SWMacroA\right ]}
=  m\uu{-{d\over 2}}\cc\hbar\uu{{d\over 2}}\cc c\ll 0\uu {\left [ \KPMacroA\right ]}\ .
\eee

\subsection{Convention-sensitivity of scheme-dependent constants in minimal subtraction}\label{ConventionSensitivityOfCounterterms}

If we redefine $c\ll 0$ in a dimension-dependent way,
\bbb
c\ll 0 \to \tilde{c}\ll 0 \equiv f(d) \times c\ll 0\ ,
\een{DDependentC0Redefinition}
then even if we take $f(d=2) = 1$
and recalculate near $d=2$ with the same ``minimal" subtraction 
(that is, subtracting just the coefficient of the pole ${1\over{d-2}}$), 
we get a different finite part of the operator dimension by terms proportional to $f\pr(d=2)$.
The $d$-dependent redefinition \rr{DDependentC0Redefinition} of
$c\ll 0$ is equivalent to redefining $\xi$ by
\bbb
\xi \to \tilde{\xi}\ = f(d)\uu{- {2\over d}} \, \xi\ ,
\eee
where we have used the relationship \rr{C0InTermsOfNishidaSonXiSummaryInAppendix} between
$\xi$ and $c\ll 0$.
So the result 
\rr{DimRegResultInXi}
for the $O(c\ll 1)$ piece of $\D\ll Q$, which goes as $\xi\uu{{{d-1}\over 2}}$, transforms as
\bbb
\D\ll Q\cc\bigg |\ll{O(c\ll 1)} \to \big [ \cc f(d) \cc \big ]\uu{- {{d-1}\over d}}
\cc \D\ll Q\cc\bigg |\ll{O(c\ll 1)}\ .
\eee
If we parametrize the bare $\D\ll Q\cc\bigg |\ll{O(c\ll 1)} $ near
$d=2$ by
\bbb
\D\ll Q\cc\bigg |\ll{O(c\ll 1)}  = {{b\ll{-1}}\over{d-2}} + b\ll 0\cc (d-2)\uu 0 + O((d-2))\ ,
\eee
then the coefficients $b\ll{-1,0}$ transform as
\bbb
b\ll{-1}\to [ \cc f(2) \cc \big ]\uu{- \hh}\cc b\ll{-1}\ ,
\xxnn
b\ll 0\to  [ \cc f(2) \cc \big ]\uu{- \hh}\cc [b\ll 0 - \hh \cc f\pr(2) b\ll {-1}]\ .
\een{SchemeDependenceFromConventionDependence}
In other words, the term $b\ll 0$ does not transform
covariantly under a change in the regularization procedure even within dimensional regularization, transforming
with an additive shift proportional to $f\pr(2) b\ll 0$.  So, even within dimensional regularization
with minimal subtraction, there is an ambiguity that 
affects the renormalized answer.

Of course, there is nothing metaphysical about scheme ambiguities: In local quantum
field theories, they always correspond to coefficients of local terms
in the Hamiltonian.  The particular scheme-dependence corresponding to the ambiguity 
\rr{SchemeDependenceFromConventionDependence} corresponds to a counterterm 
localized at the edge of the atom droplet in the harmonic trap, specifically
the edge counterterm $\d\bf(X){\bf Y}\uu{+1}$. In the present
article we have taken care to remove this ambiguity by defining
our Lagrangian couplings in all dimensions $d$; see the discussion in Sec.~\ref{ConventionsForXiDefinitionSummary}.

\end{appendix}

\section*{Ackowledgments}%

The authors thank the Burke Institute at Caltech for hospitality while this work was in progress; S.H. also thanks the Simons Center for Geometry and Physics for hospitality during the program, ``Quantum Mechanical Systems at Large Quantum Number,'' while this work was in progress. 
We are also grateful to Domenico Orlando for early discussions, and particularly for reviewing the details and results of our 
calculation of the order $Q\uu 0$ Casimir contribution of $- 0.29416$ to the ground-state operator dimension, which the authors of \cite{ToAppearReffertEtAl} have subsequently
confirmed by a different method. 
The work of S.H. is supported by the World Premier International Research
Center Initiative (\textsc{wpi} Initiative), \textsc{mext}, Japan; by the \textsc{jsps} Program for Advancing Strategic International Networks to Accelerate the Circulation of Talented Researchers;
and also supported in part by \textsc{jsps kakenhi} Grant Numbers \textsc{jp22740153, jp26400242}.



\end{document}